\documentclass[twocolumn,showpacs,preprintnumbers,amsmath,amssymb,nofootinbib]{revtex4}

\usepackage{graphicx}
\usepackage{dcolumn}
\usepackage{bm}
\usepackage{amssymb}
\usepackage{amsmath}
\usepackage{amsthm}

\begin{document}


\title{Interference Peak in the Spectrum of Bremsstrahlung on Two Amorphous Targets}

\author{M.V.~Bondarenco}
 \email{bon@kipt.kharkov.ua}
 \author{N.F.~Shul'ga}
 \affiliation{NSC Kharkov Institute of Physics and Technology, 1 Academic St.,
61108 Kharkov, Ukraine}

\date{\today}

\begin{abstract}
We investigate the interference pattern in the spectrum of
non-dipole bremsstrahlung on two amorphous foils. Apart from
suppression at lowest $\omega$, the spectrum exhibits an enhancement
adjacent to it. In classical electrodynamics, the net effect of
suppression and enhancement proves to be zero. We study the location
and the origin of the spectral features, comparing predictions of
full Moli\`{e}re averaging with those of the
Gaussian averaging with Coulomb corrections to the rms multiple scattering angle. 
Comparison with experimental data, and with previous theoretical
predictions is presented.


\end{abstract}





\maketitle

\section{Introduction}

Suppression of the low-energy part of bremsstrahlung from
ultrarelativistic electrons passing through an amorphous scattering
medium is the celebrated LPM-effect \cite{LPM-old-reviews}, named
after Landau and Pomeranchuk who predicted it qualitatively
\cite{LPM}, and Migdal \cite{Migdal} who provided an accurate theory
for bremsstrahlung in a semi-infinite uniform medium. It has also
been studied for finite-thickness targets
\cite{Ternovskii,Pafomov,Blankenbecler-Drell,Zakharov,BK-LPM}, and
for the opposite limit of thin target, which is of intrinsic
interest \cite{Ternovskii,SF}. The theoretical predictions were
accurately tested experimentally only within the last two decades
\cite{LPM-experim}, and the theoretical and experimental state of
the art is reviewed in \cite{Klein,ASF-LPM,BK-PhysRep}.

A natural further step, both for practice and for theory, is to
study composite targets. One of the simplest configurations thereof
is a sequence of thin plates with a gap between them comparable to
the photon formation length
\begin{equation}\label{l_f-def}
l_{\text{f}}=\frac{2E_e}{m^2_e}\left(\frac{E_e}{\omega}-1\right).
\end{equation}
The latter quantity depends on the electron energy $E_e$ and photon
energy $\omega$ in the window covered by the  detector, $m_e$
standing for the electron mass. The problem of structured target as
applied to the gamma-quantum emission, was first discussed by
Blankenbecler \cite{Blankenbecler,Blankenbecler-FunctionalInt}, who
treated it at the basis of the model for multiple scattering devised
in \cite{Blankenbecler-Drell}. Subsequently, other authors
\cite{Zakharov,BK-structured} extended their formalisms to handle
this problem.

An experimental verification of the existence of interference
effects in bremsstrahlung on compound targets has recently been
undertaken at CERN \cite{Thomsen,Andersen-2foils,NA63-plans}. It was
emphasized \cite{Andersen-2foils,NA63-plans} that high electron
energy allows achieving macroscopic values for the photon formation
length (\ref{l_f-def}), and thereby study its manifestations simply
by tuning the distance between the scattering plates by a micrometer
screw. But even for CERN energies, $E_e\approx 178$ GeV, the
coefficient in Eq.~(\ref{l_f-def}) amounts only
$\frac{2E_e}{m_e^2}\approx 0.4\,\mu$m, wherefore to make
$l_{\text{f}}$ practically macroscopic ($\gtrsim10^{-2}$ mm), one
needs to consider sufficiently small photon energies compared to
$E_e$, which then renders the problem essentially classical. At the
same time, condition $\omega/E_e\gg 10^{-4}$ can be safely
fulfilled, wherewith the contribution from transition radiation may
be neglected. The experiment \cite{NA63-plans} conducted at such
conditions favored the theory \cite{Blankenbecler}, but also found
some departures from it, and, surprisingly, from other theoretical
predictions, which were expected to be more accurate.

To clarify the origin of remaining discrepancies, it may be
instructive to recollect underlying assumptions for the theories at
the market. It is fair to say that Blankenbecler and Drell
\cite{Blankenbecler-Drell} in essence extended Landau and
Pomeranchuk's approach \cite{LPM} to cover the case of
finite-thickness target, and incorporate quantum effects. At that, a
simplified model of scattering in a medium was adopted, in which
transverse dimensions of the target random field exceeded that of
the electron wave packet, allowing to exactly integrate over
electron impact parameters and to reduce the photon emission
amplitude to an integral over particle trajectory, like in classical
mechanics. A strong simplification was made then, in the spirit of
\cite{LPM}, by replacing the medium average of an oscillatory
integrand by the corresponding oscillatory function of averaged
variables. That yielded the angle-integral spectrum in the form of a
double time integral, which for most cases of interest had to be
computed numerically. Along these lines, Blankenbecler
\cite{Blankenbecler} evaluated the spectrum of radiation on $N$
plates, and discovered an additional maximum, or ``shoulder" in the
region where the gap width becomes commensurable with the radiation
coherence length.

The approach of Blankenbecler and Drell has the merit of being
simple and qualitatively correct. However, it suffers from a lack of
accuracy due to the oversimplified averaging (the neglect of
fluctuations), which may be essential for comparison with
experiments. An attempt to improve it was undertaken in
Blankenbecler's later paper \cite{Blankenbecler-FunctionalInt},
where a correlation of the amplitude and the phase of the integrand
in the double time integral was taken into account, while other
fluctuations were neglected, as before. As we will show, however,
the correction so evaluated improves the accuracy only for the case
of weak scattering in the plates, whilst in the opposite case of
strong (non-dipole) scattering, in which interference effects are
actually the strongest, this correction is rather counterproductive.

Almost simultaneously with \cite{Blankenbecler-Drell,Blankenbecler},
Zakharov \cite{Zakharov} advanced with a calculation of the LPM
effect in finite targets including proper averaging. It was built
upon techniques originally developed for particle physics, notably
the impact parameter representation. The merit of the latter is that
it naturally invokes Fourier transforms of scattering-angle
distribution functions, which from solution of the kinetic equation
express as simple exponentials of the Fourier-transformed scattering
differential cross-section. The main attention in papers
\cite{Zakharov} was payed to a single finite-thickness target, while
for structured target, the results were presented only graphically.
Zakharov noticed that the interference features in his calculated
spectrum were less pronounced than those in \cite{Blankenbecler},
albeit attributed that to the crudeness of the Blankenbecler-Drell
model for scattering medium, rather than to an oversimplification of
their averaging procedure.

Shortly after, Baier and Katkov \cite{BK-structured} proposed a
technique which allowed them to handle the case of $N$ scattering
plates analytically. To this end, an alternative form of the double
time integral representation for the radiation spectrum was adopted,
and radiation spectra for $N$ plates were  calculated by
mathematical induction, assuming all the plates to be identical,
equidistant, and the scattering angle distributions in them to be
Gaussian. The influence of the transition radiation was taken into
account too, which can be valuable for lower-energy experiments.
Baier and Katkov ultimately restricted their analysis to the case of
plates infinitesimally thin compared to the gap width, and
demonstrated that instead of one maximum, the radiation spectrum
features a sequence of maxima and minima of decreasing amplitude.

The common drawback shared by all the abovementioned approaches is
that the resulting expressions for the radiation spectrum are rather
unwieldy, and relations of the spectral features with physical
parameters in the problem remain obscure. What is more serious,
however, is that it led to a number of qualitative controversies in
the literature, the most noticeable among which are the following:
\begin{itemize}
\item It was suggested in \cite{Blankenbecler} that the position of the greatest spectral maximum corresponds to a situation when the photon formation length (\ref{l_f-def}) matches the distance between the plate centres. But in that case, the resonance condition ought to involve an additional factor $2\pi$, as in coherent bremsstrahlung (see \cite{CB,NA63-plans}). However, therewith it would definitely contradict the experiment \cite{NA63-plans}.
\item In the integral expressing the radiation spectrum, the oscillating function typically has the argument $\frac{\omega t}{2\gamma^2}(1+\gamma^2\theta^2)$, where $\gamma$ is the electron's Lorentz factor, and $\theta$ the radiation angle. The use of the latter expression suggests that for the distance $l_{\text{g}}+l$ between the plate centers, the main spectral maximum should be located at
\[
\omega\simeq2\pi\frac{2\gamma^2}{(l_{\text{g}}+l)(1+\gamma^2\theta^2)},
\]
involving factor $2\pi$, as expected, but also a denominator
$1+\gamma^2\theta^2$, which might compensate it. Practically,
though, it is not a priori obvious, with respect to which direction
the radiation angles should be counted, and what are their typical
values. If one estimates them as
$\theta\sim\max\{\gamma^{-1},\sigma\}$, where $\sigma$ is the rms
scattering angle in the plates \cite{BK-structured}, then for
$\sigma\gg\gamma^{-1}$ the compensation between the $2\pi$ factor
and the denominator can be significant. But numerically, such
estimates still contradict the experimental findings
\cite{NA63-plans}.
\item Concerning the comparative strength of the contributions from the enhancement region and the suppression region at low $\omega$, it is unobvious whether their integrated effect must be zero, as suggested in \cite{Bell}, or the suppression must dominate in the total energy losses, as one might anticipate from the Migdal's theory \cite{Migdal} of radiation in a thick target, where no enhancement manifests itself in the  spectrum at all.
\end{itemize}
In the absence of cogent explanation for those controversies, the
state of the theory can hardly be regarded as satisfactory. Besides
that, it is becoming desirable further to consider the case of
plates of unequal thickness, and to fully scrutinize the role of
non-Gaussian (Coulomb) tails in scattering, insofar as approaches of
Baier and Katkov, as well as Blankenbecler and Drell rested on
Gaussian description of scattering.

Given rather fundamental significance of the present problem, it
makes sense at present to reappraise its theoretical treatment. In
this paper we shall derive physically more transparent
representations for the radiation  spectrum, which will help clarify
the matter as a whole, and at the same time improve the numerical
accuracy. To simplify the problem, we will consider the case of two
\emph{thin} scattering plates (foils), which allows avoiding double
time integral representations for the radiation spectrum. We will
also limit ourselves to classical electrodynamics, and neglect the
transition radiation, which are good approximations for the
performed experiments \cite{Thomsen,Andersen-2foils,NA63-plans}. On
the other hand, we will investigate in more detail the influence of
Coulomb corrections. To alleviate the connection between the theory
and the experiment, we introduce appropriate visibilities of the
interference pattern. We will be able to resolve the controversies
alluded above, and to correct abundant numerical errors existing in
the literature. Ultimately, we compare our predictions with the
recent experimental data, and show that there is no significant
discrepancy between the experiment and the theory.



\section{Spectrum of radiation at two scatterings}\label{sec:2scat}

In classical electrodynamics, the spectral-angular energy
distribution of radiation (with vector $\bm{n}$ marking the
direction of photon emission) expresses in a known way through the
charged particle trajectory $\bm{r}(t)$, $\bm{v}(t)=d\bm{r}/dt$
\cite{Jackson}:
\begin{equation}\label{init-t-int}
\frac{dI}{d\omega
d^2n}=\left|\frac{e}{2\pi}\int_{-\infty}^{\infty}dt
e^{i\omega\left[t-\bm{n}\cdot \bm{r}(t)\right]}\frac{d}{dt}
\frac{\bm{n}\times\bm{v}(t)}{1-\bm{n}\cdot \bm{v}(t)} \right|^2,
\end{equation}
$e$ being the electron charge. For our present study, it will
suffice to consider the simplest case when the electron experiences
2 abrupt scatterings, at instants $t_1$ and $t_2$:
\[
\bm{v}_1\underset{t_1}\to\bm{v}_2\underset{t_2}\to\bm{v}_3.
\]
Integration over the time in Eq.~(\ref{init-t-int}) trivially yields
\begin{subequations}\label{eq2}
\begin{eqnarray}
\frac{dI}{d\omega d^2n}&=&\left(\frac{e}{2\pi}\right)^2\left|
\bm{n}\times\bm{J}_{21}
+\bm{n}\times\bm{J}_{32}e^{i\Psi}\right|^2\label{expand-sum-sq-Ja}\\
&=&\left(\frac{e}{2\pi}\right)^2\Big[\left(\bm{n}\times\bm{J}_{21}\right)^2+\left(\bm{n}\times\bm{J}_{32}\right)^2\nonumber\\
&\,&\quad+2\left(\bm{n}\times\bm{J}_{21}\right)\cdot\left(\bm{n}\times
\bm{J}_{32}\right)\cos\Psi \Big],\label{expand-sum-sq-J}
\end{eqnarray}
\end{subequations}
with
\begin{equation}\label{J21-def}
\bm{J}_{21}(\bm{n})=\frac{\bm{v}_1}{1-\bm{n}\cdot\bm{v}_1}-\frac{\bm{v}_2}{1-\bm{n}\cdot\bm{v}_2},
\end{equation}
\begin{equation}\label{J32-def}
\bm{J}_{32}(\bm{n})=\frac{\bm{v}_2}{1-\bm{n}\cdot\bm{v}_2}-\frac{\bm{v}_3}{1-\bm{n}\cdot\bm{v}_3},
\end{equation}
\begin{equation}\label{Phi-def}
\Psi(\omega,\bm{n})=\omega (t_2-t_1)(1-\bm{n}\cdot\bm{v}_2).
\end{equation}
Eq.~(\ref{expand-sum-sq-Ja}) implies that if scatterings are well
separated, the spectral radiation amplitude can be viewed as a sum
of spectral amplitudes of radiation at single scatterings (with a
distance- and $\omega$-dependent phase shift $\Psi$ between them).
That may be conceived as the absence of effects of temporarily
nonequilibrium proper field on the radiation amplitude level.
However, observed in practice is the radiation intensity, which is
proportional to the amplitude square. It differs from a sum of
intensities on separate plates, engaging certain dependence on
$\omega$ through the cosine of the phase shift between the
scatterings.

An important generic property of Eq.~(\ref{expand-sum-sq-J}) is that
the $\omega$-integral of the interference term therein strictly
vanishes:
  \begin{equation*}
  \int_0^\infty d\omega \left(\frac{dI}{d\omega}-\frac{dI_1}{d\omega}-\frac{dI_2}{d\omega}\right)=0,
  \end{equation*}
by virtue of the relation
\begin{equation}\label{delta-t21}
\int_0^\infty d\omega
\cos\Psi=\pi\delta\left[t_{21}(1-\bm{n}\cdot\bm{v}_2)\right],
\end{equation}
and strictly non-zero value of the argument of the $\delta$-function
for $t_{21}=t_2-t_1\neq0$. The absence of interference in the total
radiative energy loss at successive random scatterings has a simple
physical reason: Equivalently, it can be expressed as a \emph{time}
integral of the particle acceleration squared, in which
time-separated scatterings obviously give non-interfering
contributions \cite{Bell}:
  \begin{equation*}
  \int_{-\infty}^\infty dt \frac{dI}{dt}=\sum_{k}\int_{-\infty}^\infty dt \frac{dI_k}{dt}.
  \end{equation*}
Hence, if a suppression occurs in one region of the spectrum
(typically at lowest $\omega$ where it is associated with the LPM
effect), it should be accompanied by a commensurable
\emph{enhancement} in some other region (which is natural to call
the anti-LPM effect).

The established property is undoubtedly general and must hold as
well for a number of scatterings greater than 2. But it may
contradict our experience that in a target with $\delta$-correlated
scatterings, interference effects to the spectrum are purely
suppressive, and there is seemingly no enhancement region at all
\cite{Migdal}. The contradiction is resolved if one takes into
account that in condensed matter, distances between atoms are very
short indeed, being usually far shorter than photon formation
length. Specifically, if we consider inequality
\begin{equation}\label{AA-gg-lf}
a_{\text{B}}=\frac{1}{m_e e^2}\ll l_{\text{f}}\approx
\frac{2E_e}{m_e^2}\frac{E_e}{\omega},
\end{equation}
where \emph{quantum} arguments stipulate that  $E_e/\omega>1$,
condition (\ref{AA-gg-lf}) is fulfilled already at $E_e\gtrsim
\frac{m_e}{e^2}\sim10^2$ MeV. In the latter (quite ubiquitous) case,
the radiation spectrum reaches its quantum end at $\omega=E_e$
sooner than the radiation enhancement can develop. Therefore, the
spectral enhancement has no room to develop, and the suppression
dominates both locally and globally \cite{Bell}.

One could imagine that for gaseous targets, where distances between
atoms are much greater than $a_{\text{B}}$, the situation might
change, but in the latter case, it is essential that $t_{21}$ should
be regarded as a random variable. We will return to effects of
randomization of $t_{21}$ in Sec.~\ref{sec:randomization}, and
demonstrate that the anti-LPM effect still manifests itself on a
scale of photon formation length comparable with the range of
\emph{anti}-correlation, which is of order of atomic size, again.
Meanwhile, we will concentrate on studying the case of radiation on
two scattering foils with definite and macroscopic separation.



Before we proceed, however, there is another observation worth
making. For bremsstrahlung in a sufficiently thin layer of
substance, a fair approximation is the smallness of scattering
angles (dipole limit), when $(\bm{v}_1-\bm{v}_2)^2\ll
1-v^2=\gamma^{-2}$. In that case, the structure of currents
(\ref{J21-def})--(\ref{J32-def}) appreciably simplifies. Linearizing
$\bm{J}_{21}$ in $\bm{v}_{21}=\bm{\chi}_1$ and $\bm{J}_{32}$ in
$\bm{v}_{32}=\bm{\chi}_2$, integrating over radiation angles, and
averaging (\ref{expand-sum-sq-J}) over scattering angles
$\bm{\chi}_1$, $\bm{\chi}_2$ yields
\begin{equation}\label{g}
\left\langle\frac{dI}{d\omega}\right\rangle=\left(\left\langle\frac{dI_1}{d\omega}\right\rangle+\left\langle\frac{dI_2}{d\omega}\right\rangle\right)\!
\left(1+\frac{2\left\langle \bm{\chi}_1\cdot
\bm{\chi}_2\right\rangle}{\left\langle
\bm{\chi}_1^2\right\rangle+\left\langle
\bm{\chi}_2^2\right\rangle}g_{\text{dd}}(\Omega)\right),
\end{equation}
with
\begin{equation}\label{g-dipole}
g_{\text{dd}}(\Omega)=\frac32\int_0^\infty d\Theta^2
\frac{1+\Theta^4}{(1+\Theta^2)^4}\cos{\Omega(1+\Theta^2)},
\end{equation}
$\Omega=\frac{\omega t_{21}}{2\gamma^2}$, and $\Theta=\gamma\theta$.
Since in most cases, successive scatterings are not causally
connected, the correlator in (\ref{g}) reduces to a product of mean
scattering angles:
\[
\left\langle \bm{\chi}_1\cdot \bm{\chi}_2\right\rangle=\left\langle
\bm{\chi}_1\right\rangle \cdot \left\langle
\bm{\chi}_2\right\rangle.
\]
For amorphous targets, mean deflection angles are obviously zero,
wherefore the whole interference in the spectrum in this
approximation vanishes. To preserve the interference effects between
the targets, one thus needs either to render both scatterings
asymmetric\footnote{Natural candidates for such deflectors could be
magnets, but strong magnets with dimensions smaller than
$l_{\text{f}}$ are challenging to manufacture. More compact
deflectors could be created with the aid of lasers, as
proposed in \cite{Andersen-2foils}, or bent crystals
.}, or to arrange \emph{non-dipole} conditions for radiation in the
amorphous target, which may be simpler to bring into practice, and
will be the subject of our study in what follows.


\section{Spectrum of non-dipole radiation from one
plate}\label{sec:1plate}

To make our presentation self-contained, before addressing
interference effects in a composite target, let us recapitulate the
properties of bremsstrahlung on a single amorphous foil. This
section gives the account of relevant formulae, highlighting the
differences between Gaussian and Moli\`{e}re averaging procedures,
as well as between frameworks of different authors. We will also
propose two our own recipes for incorporating Coulomb corrections,
which give a reasonable balance between precision and simplicity for
the addressed case.

Radiation spectrum at scattering to a definite angle of arbitrary
size (which is small by absolute value, but may be sizable compared
with the typical radiation angle $\gamma^{-1}$) expresses by an
integral (see, e.g., \cite{BLP})
\begin{eqnarray}\label{I1}
\frac{dI_1}{d\omega}=\left(\frac{e}{2\pi}\right)^2\int d^2 n
\left(\bm{n}\times\bm{J}_{21}\right)^2=\frac{2e^2}{\pi}F(\gamma\chi/2),
\end{eqnarray}
where
\begin{equation}\label{F-def}
F(\xi)=\frac{2\xi^2+1}{\xi\sqrt{\xi^2+1}}\ln\left(\xi+\sqrt{\xi^2+1}\right)-1.
\end{equation}
Small-angle expansion of (\ref{F-def}) can be obtained as
\begin{subequations}
\begin{eqnarray}
F(\xi)&=&-\sum^\infty_{n=1}\frac{(n-1)!(n+1)! }{(2n+1)!}(-4\xi^2)^{n}.\label{F-dipole-series}\\
&=&\frac43\xi^2-\frac45\xi^4+\ldots=\frac13(\gamma\chi)^2-\frac1{20}(\gamma\chi)^4+\ldots. \nonumber\\
\label{F-dipole}
\end{eqnarray}
\end{subequations}
Series (\ref{F-dipole-series}) has finite (actually, unit)
convergence radius, which is natural in view of the presence in
(\ref{F-def}) of the structure $\sqrt{\xi^2+1}$.

The large-angle asymptotics of (\ref{F-def}) is logarithmic:
\begin{equation}\label{F-la}
F(\xi)\underset{\xi\to\infty}\simeq2\ln 2\xi-1=2\ln\gamma\chi-1.
\end{equation}
The logarithm here stems from the fact that although for a
large-angle scattering, the radiation cones from the initial and
final electron lines are well separated, the region between them is
filled by an enhanced radiation, owing to the interference between
the cones. It is integration over this inter-jet domain which gives
rise to the large logarithm in (\ref{F-la}).

\begin{figure}
\includegraphics{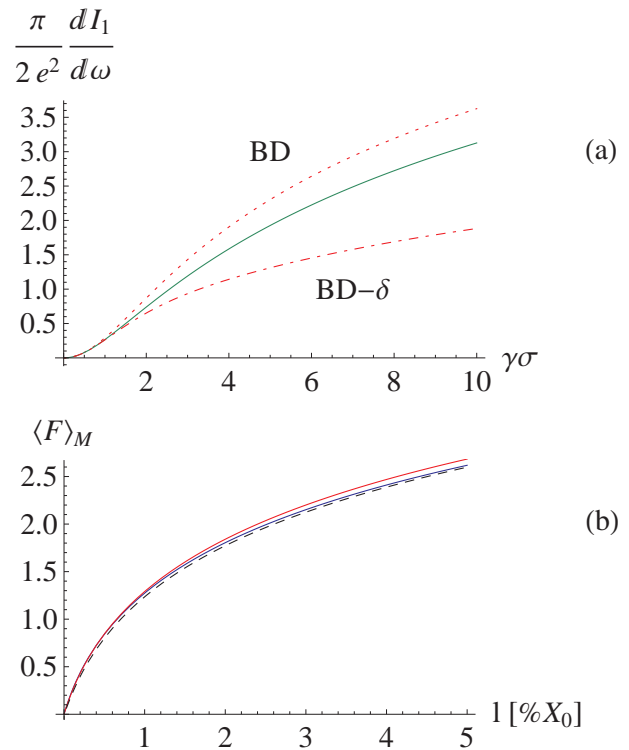}
\caption{\label{fig:F} (a) Green solid curve, $\left\langle
F\right\rangle_{\sigma}$ [Eq.~(\ref{Fsigma-change-var}) or
(\ref{K1^2Gauss})]; red dashed curve, $F$ [Eq.~(\ref{F-def}) or
Eq.~(\ref{FB(T)=F(Sigma2)})]; red dot-dashed curve, the corrected
Blankenbecler-Drell approximation
(\ref{Ftilde-Blankenbecler-thintarget}). (b) Red curve (upper);
Moli\`{e}re averaging for gold, blue curve, Moli\`{e}re averaging
for carbon; dashed curve, Gaussian averaging by
Eq.~(\ref{Highland-approx}). }
\end{figure}

\subsection{Gaussian averaging over scattering angles}

Eqs.~(\ref{I1}), (\ref{F-def}) may be used to derive radiation
spectrum in a thin layer of substance [under condition when a target
thickness is much smaller than the photon formation length
(\ref{l_f-def}) in the considered range of $\omega$]. To this end,
(\ref{I1}) must be averaged with the appropriate distribution in
scattering angles in the target \cite{Ternovskii,SF}. In the
simplest approximation, this distribution can be taken to be
Gaussian. Averaging over scattering angles with the Gaussian
distribution with rms scattering angle $\sigma$,
\begin{equation}\label{Gauss-distr}
\frac{dw_{\sigma}}{d\chi^2}=\frac1{\sigma^2}e^{-\chi^2/\sigma^2},
\end{equation}
one gets
\begin{equation}\label{I1sigma}
\left\langle\frac{dI_1}{d\omega}\right\rangle_{\sigma}=\frac{2e^2}{\pi}\left\langle
F\right\rangle_{\sigma},
\end{equation}
where
\begin{eqnarray}\label{Fsigma}
\left\langle F\right\rangle_{\sigma}&=&\int_0^{\infty}d\chi^2\frac1{\sigma^2}e^{-\chi^2/\sigma^2} F(\gamma\chi/2)\nonumber\\
&=&\frac{8}{\gamma^2\sigma^2}\int_0^{\infty}d\xi\xi
e^{-\frac{4\xi^2}{\gamma^2\sigma^2}} F(\xi).
\end{eqnarray}
Inserting here (\ref{F-dipole-series}), and integrating termwise
yields
\begin{eqnarray}\label{F-aver-series}
\left\langle
F\right\rangle_{\sigma}=-\sum^{\infty}_{n=1}\frac{(n-1)!n!(n+1)!}{(2n+1)!}\left(-\Sigma^2\right)^n,
\end{eqnarray}
with $\Sigma=\gamma\sigma$. Series (\ref{F-aver-series}) diverges
for any finite $\Sigma$ (not surprisingly since it was derived by
integrating a series beyond its convergence domain); nonetheless,
the sequence of its terms may be used as an asymptotic expansion.

The first two terms in Eq.~(\ref{F-aver-series}) are
\begin{equation}\label{Fsigma-dipole}
\left\langle F\right\rangle_{\sigma} \underset{\Sigma\ll1}\simeq
\frac13\Sigma^2-\frac1{10}\Sigma^4.
\end{equation}
Note that the coefficient at  the leading-order term is the same as
in (\ref{F-dipole}), but the coefficient at the next-to-leading
order term is twice greater.

To obtain large-$\Sigma$ asymptotics of $\left\langle
F\right\rangle_{\sigma}$, it is advantageous to change in
(\ref{Fsigma}) the integration variable to $\xi=\sinh\frac{w}2$:
\begin{eqnarray}\label{Fsigma-change-var}
\left\langle F\right\rangle_{\sigma}
&=&\frac2{\Sigma^2}e^{\frac{2}{\Sigma^2}} \int_0^{\infty}dw e^{-\frac{2}{\Sigma^2}\cosh w} w\cosh w-1\nonumber\\
&=&e^{\frac{2}{\Sigma^2}}K_0\left(\frac{2}{\Sigma^2}\right)-1\nonumber\\
&\,&\quad+\frac2{\Sigma^2}e^{\frac{2}{\Sigma^2}} \int_0^{\infty}dw
we^{-w-\frac{2}{\Sigma^2}\cosh w}.
\end{eqnarray}
The remaining integral in (\ref{Fsigma-change-var}) is a bounded
function, ranging from 0 to 1. At large $\Sigma$, this term vanishes
due to the pre-factor $\Sigma^{-2}$, so the corresponding
asymptotics of $\left\langle F\right\rangle_{\sigma}$ is determined
by that of Macdonald function at the origin: $K_0(z)\underset{z\to
0}{\simeq} \ln\frac2z-\gamma_{\text{E}}$. Employing this in
(\ref{Fsigma-change-var}), we get \cite{SF}
\begin{equation}\label{Fsigma-la}
\left\langle F\right\rangle_{\sigma} \underset{\Sigma\gg1}\simeq
2\ln\Sigma-\gamma_{\text{E}}-1.
\end{equation}

\subsection{Blankenbecler's approximations}

It is instructive to compare result (\ref{I1sigma})--(\ref{Fsigma})
with representations derived within frameworks of Blankenbecler and
Drell \cite{Blankenbecler}, and Baier and Katkov
\cite{BK-structured}, under the same conditions of (geometrically)
thin target and Gaussian scattering.

Blankenbecler's result [see \cite{Blankenbecler}, Eq.~(30)] for the
thin-target limit reads
\begin{subequations}\label{F-Blankenbecler-thintarget}
\begin{equation}\label{F-Blankenbecler-thintarget-int}
F_{\text{B}}(T)=\int_0^1 dw\left(\frac{1+3T}{1+6Tw(1-w)}-1\right).
\end{equation}
Evaluation of this integral yields
\begin{equation}\label{}
F_{\text{B}}(T)=F\left(\sqrt{3T/2}\right),
\end{equation}
\end{subequations}
where $F$ is given by Eq.~(\ref{F-def}). Hence, the correspondence
between $T$ and $\Sigma$ is
\begin{equation}\label{T=Sigma2/6}
T=\Sigma^2/6,
\end{equation}
and the approximation of Blankenbecler consists in substituting the
rms scattering angle to the non-averaged (fixed-scattering-angle)
radiation spectrum:
\begin{equation}\label{FB(T)=F(Sigma2)}
F_{\text{B}}(T)=F(\Sigma/2).
\end{equation}
This correspondence is what one could expect, given that
Blankenbecler and Drell \cite{Blankenbecler-Drell}, similarly to
Landau and Pomeranchuk \cite{LPM}, replaced an average of an
oscillatory function in the integrand of a double time integral by
the oscillatory function with average amplitude and phase. As is
seen from comparison of asymptotics (\ref{Fsigma-la}) and
(\ref{F-la}), however, such a simple-minded replacement deviates
from the exact result only by $\sim 15\%$.

Yet another possible source of inaccuracy, though, may stem from the
fact that for numerical calculations, Blankenbecler adopted relation
\begin{equation}\label{Blankenbecler-Q2def}
\left\langle Q^2_{\perp}\right\rangle=\frac{2\pi
m_e^2}{e^2}\frac{l}{X_0}
\end{equation}
(with $l$ the target thickness and $X_0$ the radiation length). The
coefficient in (\ref{Blankenbecler-Q2def}) differs by a constant
factor $1/2$ from that in the Rossi formula
\begin{equation}\label{Rossi}
\left\langle Q^2_{\perp}\right\rangle=\frac{4\pi
m_e^2}{e^2}\frac{l}{X_0}.
\end{equation}
Accordingly, Blankenbecler defined the scaled target thickness [see
\cite{Blankenbecler-Drell}, Eq.~(10.1), \cite{Blankenbecler},
Eq.~(16)]
\[
T=\frac{\pi}{3\alpha}\frac{l}{X_0},
\]
consistent with the correspondence rule (\ref{T=Sigma2/6}). But
physically, the coefficient in the relation between $\left\langle
Q^2_{\perp}\right\rangle$ and ${l}/{X_0}$ should vary
logarithmically with $l$. More accurate description of multiple
Coulomb scattering will be furnished in
Sec.~\ref{subsec:1plateMoliere}.

In \cite{Blankenbecler-FunctionalInt}, Blankenbecler derived a
correction to Eq.~(\ref{F-Blankenbecler-thintarget-int}) [see
\cite{Blankenbecler-FunctionalInt}, Eq.~(109)]:
\begin{eqnarray}\label{Ftilde-Blankenbecler-thintarget}
\tilde F_{\text{B}}(T)=\int_0^1
dw\Bigg(\frac{1+\frac32T}{1+6Tw(1-w)}\qquad\qquad\nonumber\\
+\frac{\frac32T}{\left[1+6Tw(1-w)\right]^2}-1\Bigg).
\end{eqnarray}
At small $T$, i.e., small $\Sigma$, it rather neatly reproduces the
next-to-leading order term in the exact small-angle expansion
(\ref{Fsigma-dipole}), provided one uses here the correspondence
(\ref{T=Sigma2/6}):
\begin{equation}\label{Ftilde-Blankenbecler-thintarget-smallSigma}
\tilde F_{\text{B}}(T)\underset{\Sigma\ll1}\simeq
2T-\frac{33}{10}T^2= \frac13\Sigma^2-\frac{11}{120}\Sigma^4.
\end{equation}
However, at large $\Sigma$ (or $T$), approximation
(\ref{Ftilde-Blankenbecler-thintarget}) fails by a factor of 2 [see
Fig.~\ref{fig:F}(b), red dot-dashed curve], and can not be regarded
as tenable there.

\subsection{Baier and Katkov's representation}

Baier and Katkov's result for the case of one foil
\cite{BK-structured} reads\footnote{We restored factor $x^{-1}$
missing in \cite{BK-structured}.}
\begin{equation}\label{F-BK-1target}
F_{\text{BK}}(b)=\int_0^{\infty}\frac{dx}x e^{-x}\int_0^1
dw\left\{1-\frac1{\left[1+\frac{x}{b}  w(1-w)\right]^2}\right\}.
\end{equation}
Evaluation of this integral re-obtains expression (\ref{Fsigma}),
once one identifies
\begin{equation}\label{b-Sigma2}
b=\Sigma^{-2},
\end{equation}
and
\begin{equation*}\label{}
F_{\text{BK}}(\Sigma^{-2})=\left\langle F\right\rangle_{\sigma}.
\end{equation*}
Baier and Katkov stated their own prescription for computation of
$b^{-1}$, which will be compared with our prescription in the next
subsection.

\subsection{Weighting with the Moli\`{e}re distribution. Impact parameter
representation}\label{subsec:1plateMoliere}

For very thin targets, the distribution in scattering angles can
significantly deviate from a Gaussian, while in the extreme $l\to0$,
it must turn proportional to the differential cross-section of
single scattering. To be more accurate, one may utilize the general
(Moli\`{e}re) solution of the kinetic equation in terms of a
Fourier-Bessel integral \cite{Bethe}:
\begin{equation}\label{Moliere-distr-def}
\frac{dw_{\text{M}}}{d^2\chi}=\frac1{(2\pi)^2}\int d^2 r
e^{i\bm{r}\cdot\bm{\chi}} e^{-nl \int d\sigma(\chi)[1-J_0(r\chi)]},
\end{equation}
where $d\sigma(\chi)$ is the scattering differential cross-section.
It may be noted that at substantial thicknesses,
(\ref{Moliere-distr-def}) is approximable by a Gaussian plus
corrections \cite{Bethe}, but we will not resort to such
approximations here, dealing directly with the exact integral
representation (\ref{Moliere-distr-def}).

According to Eqs.~(\ref{expand-sum-sq-J}), (\ref{J21-def}),
\begin{eqnarray}\label{80}
\frac{dI}{d\omega }&=&\left(\frac{e}{2\pi}\right)^2\int d^2n\left(
\frac{\bm{n}\times\bm{v}_1}{1-\bm{n}\cdot\bm{v}_1}-\frac{\bm{n}\times\bm{v}_2}{1-\bm{n}\cdot\bm{v}_2}
\right)^2\nonumber\\
&=&\left(\frac{e}{\pi}\right)^2\int d^2n\Bigg( \frac{\bm{n}-\bm{v}_1}{\gamma^{-2}+(\bm{n}-\bm{v}_1)^2}\nonumber\\
&\,&\qquad\qquad\qquad-\frac{\bm{n}-\bm{v}_2}{\gamma^{-2}+(\bm{n}-\bm{v}_2)^2}
\Bigg)^2,
\end{eqnarray}
where the last integral effectively extends over the transverse
plane of small angle differences. Since the two terms in parentheses
in (\ref{80}) only differ by a shift in the $\bm{n}_{\perp}$ plane,
it may be expedient to expand them into Fourier integrals, so that
the angular shift converts to a phase factor:
 \begin{equation}\label{wf-angle-wf-imp-par}
\frac{\bm{n}-\bm{v}}{\gamma^{-2}+(\bm{n}-\bm{v})^2}=\frac{i}{2\pi}\int
d^2 r e^{i(\bm{n}-\bm{v})\cdot \bm{r}} \frac{\partial}{\partial
\bm{r}}K_0(r/\gamma).
\end{equation}
Considering that $\omega(\bm{n}-\bm{v})$ is the photon's transverse
momentum, $\bm{r}/\omega$ can be interpreted as the photon's impact
parameter relative to the parent electron.

Inserting (\ref{wf-angle-wf-imp-par}) to Eq.~(\ref{80}) obtains
\begin{subequations}
\begin{eqnarray}
\frac{dI}{d\omega } &=&\left(\frac{e}{\pi}\right)^2\int d^2r
\left[\frac{\partial}{\partial
\bm{r}}K_0(r/\gamma)\right]^2\left|1-e^{i(\bm{v}_1-\bm{v}_2)\cdot
\bm{r}}
\right|^2\nonumber\\
&\,&\label{40a}\\
&=&\frac{4e^2}{\pi\gamma^2}\int_0^{\infty} dr r
K_1^2(r/\gamma)\left[1-J_0\left(\chi
r\right)\right].\label{K1^2(1-J)}
\end{eqnarray}
\end{subequations}
It  can be checked that evaluation of the integral in
(\ref{K1^2(1-J)}) reproduces the explicit form (\ref{I1}),
(\ref{F-def}).

The reader familiar with the light-cone formalism
\cite{light-cone-wf} will recognize
$\frac{e}{\pi}\frac{\partial}{\partial \bm{r}}K_0(r/\gamma)$ as
being (the low-$\omega$ approximation for) the electron-photon
component of the electron wave function, with the vector index
accounting for the photon polarization. In this spirit,
$\frac{e}{\pi}\frac{\partial}{\partial
\bm{r}}K_0(r/\gamma)\left(1-e^{i(\bm{v}_1-\bm{v}_2)\cdot
\bm{r}}\right)$ amounts the complete wave function after electron
scattering to a definite angle, and for $\bm{v}_2\neq\bm{v}_1$ this
wave function differs from zero. Correspondingly, the integral of
its modulus square gives the total probability of emission of a
photon with energy $\omega$, i.e., the radiation spectrum.
Interpretation of this kind may be noteworthy for intuitive
understanding of the final results, albeit in this paper we will
refrain from using it technically, and adhere to the standard
formulation of classical electrodynamics.\footnote{This attitude
also saves us from encountering divergences, compared with
\cite{Blankenbecler-Drell,Zakharov,Blankenbecler,BK-structured}
where subtraction of  divergent `vacuum' terms from the final result
was needed.}

To proceed, when expression (\ref{K1^2(1-J)}) is  averaged over
scattering angles with a Gaussian distribution (\ref{Gauss-distr}),
it becomes
\begin{eqnarray}\label{K1^2Gauss}
\left\langle\frac{dI}{d\omega}\right\rangle_{\sigma}
&=&\int d^2\chi \frac{dw_{\sigma}}{d^2\chi}\frac{dI}{d\omega}\nonumber\\
&=&\frac{4e^2}{\pi\gamma^2}\int_0^{\infty} dr r
K_1^2(r/\gamma)\left(1-e^{-\sigma^2r^2/4}\right).
\end{eqnarray}
On the other hand, averaging with Moli\`{e}re distribution
(\ref{Moliere-distr-def}) leads to the form
\begin{equation}\label{K1^2(1-edsigma)}
\left\langle\frac{dI}{d\omega}\right\rangle_{\!\text{M}}
=\frac{4e^2}{\pi\gamma^2}\int_0^{\infty}\! dr r
K_1^2(r/\gamma)\!\left\{1-e^{-nl \int
d\sigma(\chi)[1-J_0(r\chi)]}\right\},
\end{equation}
essentially coinciding with the result of Zakharov \cite{Zakharov}
(see also \cite{BK-LPM}). It is also in the spirit of Glauber form
\cite{Glauber} for scattering of a high-energy composite quantum
system. (It is not crucial that in our case the scattering is
multiple, since it conserves the impact parameter, anyway.)


To use formula (\ref{K1^2(1-edsigma)}), we need further to specify
the differential cross-section of scattering on one atom. The
simplest model thereof is
\begin{equation}\label{dsigma-model}
nl d\sigma(\chi)=2\chi^2_c \frac{\chi
d\chi}{\left(\chi^2+\chi^2_1\right)^2},
\end{equation}
where $\chi^2_c=\frac{4\pi nl Z^2e^4}{p^2v^2}$, and
$\chi_1=(pR)^{-1}$, with $R$ playing the role of the atomic
screening radius. Inserting (\ref{dsigma-model}) to
(\ref{K1^2(1-edsigma)}) gives
\begin{equation}\label{exp-K1}
\left\langle\frac{dI}{d\omega}\right\rangle_{\!\text{M}}
=\frac{4e^2}{\pi}\int_0^{\infty} \!d\rho \rho
K_1^2(\rho)\!\left\{1-e^{-\frac{\chi^2_c}{\chi^2_1}\left[1-\rho\gamma\chi_1
K_1\left(\rho\gamma\chi_1\right)\right]}\right\}\!.
\end{equation}
At practice, parameter $\gamma\chi_1=(m_eR)^{-1}\ll 1$, so for any
value of $\frac{\chi^2_c}{\chi^2_1}$, for description of radiation
it is always legitimate to expand $K_1\left(\rho\gamma\chi_1\right)$
in vicinity of the origin: $1-zK_1(z)\simeq
\frac{z^2}2\left(\ln\frac{2}{z}+\frac12-\gamma_{\text{E}}\right)$.

To determine the precise value of $\chi_1$, one may note that in the
limit of small $l$,
\begin{subequations}
\begin{eqnarray}\label{rho3K12ln}
\left\langle\frac{dI}{d\omega}\right\rangle_{\text{M}}
&\underset{l\to0}\simeq&\frac{2e^2\gamma^2\chi^2_c}{\pi}\int_0^{\infty}
d\rho \rho^3
K_1^2(\rho)\nonumber\\
&\,&\qquad\qquad\times\left(\ln\frac{2}{\rho\gamma\chi_1}+\frac12-\gamma_{\text{E}}\right)\\
&=&\frac{4e^2\gamma^2\chi^2_c}{3\pi}\left(\ln\frac1{\gamma\chi_1}+\frac7{12}\right).
\end{eqnarray}
\end{subequations}
On the other hand, by definition of the radiation length (in the
approximation of complete screening), that must
equal\footnote{Corrections to the complete screening approximation
might be taken into account via a re-definition of $X_0$ in
Eq.~(\ref{43X0}).}
\begin{equation}\label{43X0}
\left\langle\frac{dI}{d\omega}\right\rangle_{\text{M}}
\underset{l\to0}\simeq\frac{4l}{3X_0}.
\end{equation}
Parameter $\chi_1$ can thus be related to phenomenological
parameters $X_0$ and $n$ via
\begin{equation}\label{X0-rho1}
\gamma\chi_1=\exp\left(\frac7{12}-\frac{m_e^2}{4 n Z^2e^6
X_0}\right).
\end{equation}

In this paper, our approach will be simply to employ relation
(\ref{X0-rho1}) in Eq.~(\ref{exp-K1}). The factor
$\frac{\chi^2_c}{\chi^2_1}$ in the exponent can be expressed in
terms of the ratio $l/X_0$:
\begin{eqnarray}\label{chi2c/chi21}
\frac{\chi^2_c}{\chi^2_1}&=&\frac{4\pi nl Z^2e^4}{m_e^2\gamma^2\chi^2_1}=\frac{\pi l}{e^2\gamma^2\chi^2_1}\frac{4 n Z^2e^6}{m_e^2}\nonumber\\
&=&\frac{\pi
}{e^2\gamma^2\chi^2_1\left(\ln\frac1{\gamma\chi_1}+\frac7{12}\right)}\frac{l}{X_0}.
\end{eqnarray}
Therewith, we have only one adjustable parameter $\gamma\chi_1$
(instead of two, for radiation with and without atom ionization or
excitation). That parameter depends on the radiation length, which
may include all the molecular binding and crystal structure effects.
At practice, though, $X_0$ itself is often inferred from
calculations by formulas for bremsstrahlung on a free atom,
neglecting the inter-atomic effects \cite{Klein,PDG,Tsai}.

Utilizing the commonly used (calculated \cite{Tsai}) values for the
radiation length, for gold we obtain $\gamma\chi_1=0.048$,
$\frac{\chi^2_c}{\chi^2_1}=5\cdot 10^4\frac{l}{X_0}$, whereas for
carbon, $\gamma\chi_1=0.0064$, $\frac{\chi^2_c}{\chi^2_1}=1.9\cdot
10^6\frac{l}{X_0}$. Note that if the screening radius is estimated
from relation $R=\frac1{m_e\gamma\chi_1}$, for gold it amounts
$0.15a_{\text{B}}$, while for carbon, $1.14a_{\text{B}}$. The
presented estimates for $R$ are generally greater than those
following from the prescription used by Zakharov
\cite{Zakharov,Tsai}
\begin{equation}\label{Zakh-a}
\ln\frac{R}{a_{\text{B}}}=\ln\frac{a_{\text{el}}}{a_{\text{B}}}+\frac1Z
\ln\frac{a_{\text{in}}}{a_{\text{B}}}=\ln(0.83 Z^{-1/3})+\frac1Z
\ln(5.2 Z^{-2/3}).
\end{equation}
For gold, Eq.~(\ref{Zakh-a}) gives $R=0.19a_{\text{B}}$, while for
carbon, $R=0.49 a_{\text{B}}$, the latter deviating distinctly from
our above estimate. Fortunately, though, the sensitivity to this
parameter is only logarithmic.

Fig.~\ref{fig:F}(b) overlays the predictions of Eq.~(\ref{exp-K1})
for carbon and for gold. Along with those, we plot formfactor
(\ref{Fsigma}) for Gaussian averaging with the rms scattering angle
evaluated by the widely used empirical formula\footnote{More
elaborate empirical formulas taking into account $Z$-dependence of
the coefficient were obtained \cite{Lynch-Dahl}, but to keep the
treatment simple, and expose numerical differences between the
approaches, we will confine ourselves here to
parameterization~(\ref{Highland-approx}).} \cite{Highland,PDG}
\begin{equation}\label{Highland-approx}
\gamma\sigma=\frac{\text{13.6
MeV}}{m_e}\sqrt{\frac{2l}{X_0}}\left(1+0.038\ln\frac{l}{X_0}\right).
\end{equation}
Formula (\ref{Highland-approx}) is known to work with an accuracy
better than 10\% for $l>10^{-3}X_0$. From Fig.~\ref{fig:F}(b) we
conclude that for bremsstrahlung on one plate, the agreement is good
in the whole range of $l$, including $l<10^{-3}X_0$. That is natural
since it can be shown that under some approximations, formula
(\ref{Highland-approx}) can actually be derived from the rigorous
representation (\ref{exp-K1}) (see Appendix \ref{appendix}).

\begin{figure}
\includegraphics{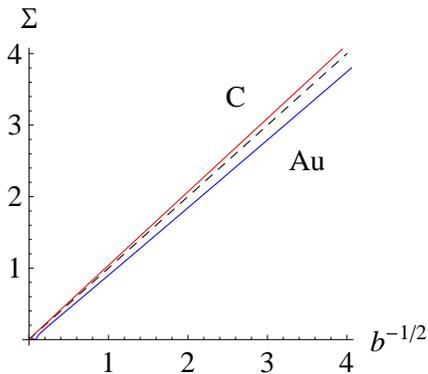}
 \caption{\label{fig:rms} Correspondence between values of rms scattering angle calculated by our Eq.~(\ref{51}) and by Baier and Katkov's Eq.~(\ref{BK-rms-prescription}). Red curve -- for carbon; blue curve -- for gold. The auxiliary dashed line marks the diagonal $\Sigma=b^{-1/2}$.}
\end{figure}

Finally, our Eq.~(\ref{51}) approximately accounting for Coulomb
corrections in multiple scattering can also be compared with the
recipe of Baier and Katkov \cite{BK-structured}:
\begin{eqnarray}\label{BK-rms-prescription}
b^{-1}=\gamma^2\chi^2_c\Bigg[2\ln(183Z^{-1/3})\qquad\qquad\qquad\qquad\quad\nonumber\\
-2(Z\alpha)^2\sum_{k=1}^{\infty}\frac1{k\left(k^2+(Z\alpha)^2\right)}+\ln
b^{-1}\Bigg].
\end{eqnarray}
Predictions of Eq.~(\ref{51}) and (\ref{BK-rms-prescription}) are
confronted in Fig.~\ref{fig:rms}, and are reasonably close, although
some differences are perceptible.

\section{Two foils}

We are now in a position to address the case of two plates. In
Eq.~(\ref{expand-sum-sq-J}), the first two terms are individual
contributions from each plate, described by formulas of
Sec.~\ref{sec:1plate}, while the nontrivial third term of
Eq.~(\ref{expand-sum-sq-J}) involves the product of currents induced
by scattering in different plates. Hence, averaging of those
currents over corresponding scattering angles proceeds
independently:
\begin{equation}\label{av-prod=prod-av}
\left\langle
\left(\bm{n}\times\bm{J}_{21}\right)\cdot\left(\bm{n}\times
\bm{J}_{32}\right)\right\rangle =\left(\bm{n}\times
\left\langle\bm{J}_{21}\right\rangle\right)\cdot\left(\bm{n}\times
\left\langle\bm{J}_{32}\right\rangle \right).
\end{equation}
As was pointed out in the end of Sec.~\ref{sec:2scat}, product
(\ref{av-prod=prod-av}) receives a suppression after integration
over azimuthal directions of $\bm{n}$. Therefore, it is reasonable
to carry out the averaging in two steps: First, to average over
azimuths of $\bm{n}$, which does not need a reference to a specific
form of the (axially-symmetric) scattering angle distribution
function. At the second step, one implements convolutions with
specific angular distribution functions.

\subsection{Azimuthal averaging for the interference term}

Granted that the currents entering Eq.~(\ref{eq2}) have similar
structure, it suffices to consider a generic form
\begin{equation}\label{J-def}
\bm{J}=\frac{\bm{v}_2}{1-\bm{n}\cdot\bm{v}_2}-\frac{\bm{v}}{1-\bm{n}\cdot\bm{v}},
\end{equation}
where $\bm{v}$ may equal $\bm{v}_1$ or $\bm{v}_3$. Averaging of
expression (\ref{J-def}) over azimuths of the scattering angle, say,
wrt direction $\bm{v}_2$ (it makes no difference since ultimately
all directions are to be integrated over) proceeds as follows.
Decomposing $\bm{v}$ into components parallel and orthogonal to
$\bm{v}_2$,
\begin{equation}\label{v-decomposition}
\bm{v}=\bm{v}_{\parallel}+\bm{v}_{\perp},
\end{equation}
with $\bm{v}_{\perp}\perp \bm{v}_2\parallel \bm{v}_{\parallel}$, we
compute the azimuthal average:
\begin{eqnarray}\label{}
\left\langle\frac{\bm{v}_{\parallel}+\bm{v}_{\perp}}{1-n_{\parallel}v_{\parallel}-\bm{n}_{\perp}\cdot\bm{v}_{\perp}}\right\rangle_{\text{azim }\bm{v}_{\perp}}\qquad\qquad\qquad\qquad\nonumber\\
=\frac1{2\pi}\int_{-\pi}^{\pi}d\phi \frac{\bm{v}_{\parallel}+\frac{\bm{n}_{\perp}}{n_{\perp}}v_{\perp}\cos\phi}{1-n_{\parallel}v_{\parallel}-n_{\perp}v_{\perp}\cos\phi}\nonumber\\
=
\frac{\bm{v}_{\parallel}+\frac{\bm{n}_{\perp}}{n^2_{\perp}}(1-n_{\parallel}v_{\parallel})}{\sqrt{(1-n_{\parallel}v_{\parallel})^2-n^2_{\perp}v^2_{\perp}}}-\frac{\bm{n}_{\perp}}{n^2_{\perp}}.\qquad
\end{eqnarray}
Now we can insert
$\bm{n}_{\perp}=\bm{n}-\bm{n}_{\parallel}=\bm{n}-\bm{v}_2\frac{n_{\parallel}}{v_2}$,
and omit the component parallel to $\bm{n}$, which does not
contribute to the vector product with $\bm{n}$ appearing in
Eq.~(\ref{av-prod=prod-av}). As a result, the azimuthally averaged
radiation amplitude assumes the form
\begin{equation}\label{approx-aver-J}
\bm{n}\times\left\langle\bm{J}\right\rangle_{\text{azim
}\bm{v}_{\perp}}=\frac{\bm{n}\times\bm{v}_2}{v_2}4\gamma^2
G(\Sigma,X),
\end{equation}
with
\begin{equation*}\label{}
4\gamma^2G=\frac{1}{1-n_{\parallel}v_2}-\frac1{\bm{n}^2_{\perp}}\left(n_{\parallel}-\frac{n_{\parallel}-v_{\parallel}}{\sqrt{(1-n_{\parallel}v_{\parallel})^2-n^2_{\perp}v^2_{\perp}}}\right).
\end{equation*}
In the small-angle approximation, $n_{\perp}=\theta\ll1$,
$v_{\perp}=\chi\ll1$, $n_{\parallel}=1-\theta^2/2$,
$v_{\parallel}=1-(\gamma^{-2}+\chi^2)/2$, we obtain
\begin{eqnarray}\label{J-scalar-def}
G&\approx& \frac{1}{2(1+\gamma^{2}\theta^2)}\nonumber\\
&\,&-\frac{1}{
4\gamma^{2}\theta^2}\!\left(\!1\!+\!\frac{\theta^2-\chi^2-\gamma^{-2}}{\sqrt{[\gamma^{-2}+(\theta-\chi)^2][\gamma^{-2}+(\theta+\chi)^2]}}\right)\nonumber\\
&\equiv&\frac{1}{2(1+\Theta^2)}-\frac{1}{
4\Theta^2}\!\left(\!1\!+\!\frac{\Theta^2-X-1}{\sqrt{(X+1-\Theta^2)^2+4\Theta^2}}\!\right),\quad
\end{eqnarray}
$X=\gamma^2\chi^2$. Function (\ref{J-scalar-def}) is everywhere
positive. At small $\Theta$, it expands as
\begin{eqnarray}\label{G(0)}
G&=&\frac12\sum_{n=0}^{\infty}\Theta^{2n}\Bigg\{(-1)^n\nonumber\\
&\,&+\frac1{2(X+1)^{n+1}}\!\left[P_{n+1}\left(\frac{X-1}{X+1}\right)- P_n\left(\frac{X-1}{X+1}\right)\right]\!\!\Bigg\}\nonumber\\
&\underset{\Theta\ll
1}\simeq&\frac12-\frac1{2(1+X)^2}+\mathcal{O}(\Theta^2),
\end{eqnarray}
where $P_n$ are Legendre polynomials \cite{Abr-Stegun} arising as
coefficients in the expansion $\left(1-2\zeta
h+h^2\right)^{-1/2}=\sum_{n=0}^\infty P_n(\zeta) h^n$.

Before proceeding to averaging over moduli of scattering angles,
i.e., $X$, it is also instructive to examine the asymptotics of $G$
as a function of $X$. At small $X$,
\begin{subequations}
\begin{eqnarray}
G\!\!&=&\!\!\frac{1}{4\Theta^2}\sum_{n=0}^{\infty}\frac{X^{n+1}}{(\Theta^2+1)^{n+1}}\Bigg[P_n\left(\frac{\Theta^2-1}{\Theta^2+1}\right)\nonumber\\
&\,&\qquad\qquad\quad
-\frac{\Theta^2-1}{\Theta^2+1}P_{n+1}\left(\frac{\Theta^2-1}{\Theta^2+1}\right)\Bigg]\,\,
\label{J-quadrupole-approx-a}\\
&\underset{X\ll1}\simeq&\frac{1}{(1+\Theta^2)^3}X+\mathcal{O}(X^2).\label{J-quadrupole-approx}
\end{eqnarray}
\end{subequations}
Here the leading term is proportional to $X\propto\chi^2$, whereas
in the dipole approximation the current is $\propto\chi$, but the
latter contribution vanishes after azimuthal averaging [see
Eq.~(\ref{g})]. Thus, the leading term (\ref{J-quadrupole-approx})
corresponds to a ``quadrupole" approximation.

On the other hand, at large $\chi$ and fixed $\theta$,
\begin{equation}\label{J-largeangle-approx}
G\underset{X\gg1}\approx\Bigg\{
\begin{array}{c}
                                                                          \frac{1}{2(1+\Theta^2)}\quad (\theta<\chi) \\
                                                                          \quad 0\qquad\quad (\theta>\chi)
                                                                        \end{array}
.
\end{equation}
Factor $\frac{1}{2(1+\Theta^2)}$ here describes the radiation from
the $\bm{v}_2$ half-line alone. As for the radiation from the
$\bm{v}$ half-line, it is smeared by the azimuthal averaging, but in
any case, the inter-jet region extends only out to polar angle
$\theta\approx\chi$. It is noteworthy that function $G$ features no
enhancement around $\theta=\chi$, which could be expected due to the
presence of the $\bm{v}$-jet. So, after the azimuthal averaging, the
latter jet does not manifests itself as a peak. For $\theta>\chi$,
the radiation is suppressed because  radiation amplitudes from
$\bm{v}_2$ and $\bm{v}$ lines strongly cancel.

\begin{figure}
\includegraphics{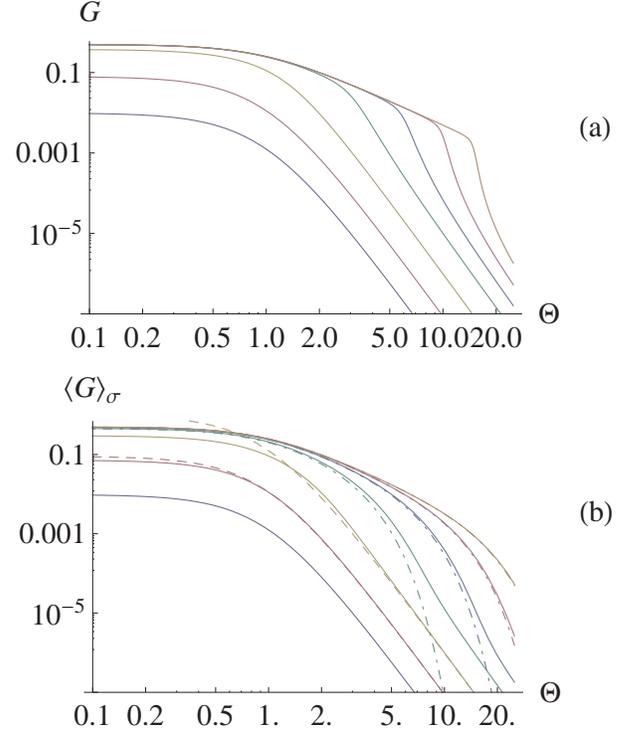}
 \caption{\label{fig:J} (a) LogLog plot of the azimuthally averaged electromagnetic current at one scattering, Eq.~(\ref{J-scalar-def}), for scattering angles $\gamma\chi=0.1$, 0.3, 1, 3, 6, 10, 15 (solid curves, bottom to top). Dashed curves, approximation (\ref{J-quadrupole-approx}); (b) Same for the Gaussian-weighted azimuthally averaged electromagnetic current at one scattering, Eq.~(\ref{J-scalar-av-chi}). Dashed curves, approximation (\ref{G-small-Sigma}). Dot-dashed curves, approximation (\ref{Gsigma-la}).}
\end{figure}

\subsection{The aggregate spectrum}\label{subsec:Aggr-spectr}

\begin{figure}
\includegraphics{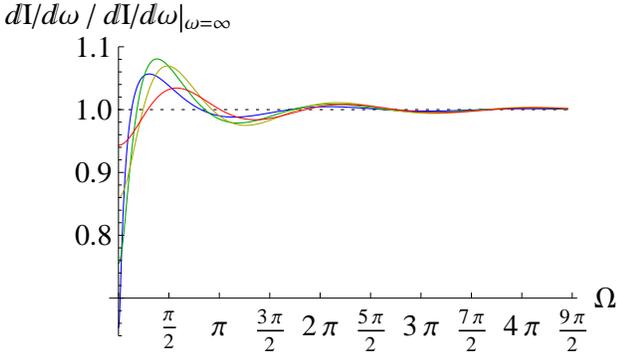}
 \caption{\label{fig:Exact-oscill} Oscillations of spectrum of the bremsstrahlung on two foils, computed for Gaussian scattering-angle distribution functions with rms $\Sigma_1=\Sigma_2=0.5$ (thin red curve), 1 (yellow), 2 (green), 5 (blue). At $\Omega\gtrsim\frac{t_{21}}{\max\{l_1,l_2\}}$, the spectrum must rise again, and reach the Bethe-Heitler constant value (see Sec.~\ref{sec:plate-formfactor}).}
\end{figure}

Employing representation Eq.~(\ref{approx-aver-J}) in
Eq.~(\ref{av-prod=prod-av}), we have
\begin{equation}\label{interf-JJ}
\left(\bm{n}\times
\left\langle\bm{J}_{21}\right\rangle\right)\cdot\left(\bm{n}\times
\left\langle\bm{J}_{32}\right\rangle \right)=-16\gamma^4\theta^2
\left\langle G\right\rangle_1\left\langle G\right\rangle_2,
\end{equation}
where
\begin{equation}\label{av-G}
\left\langle G\right\rangle_i=\int_0^{\infty}
d\chi^2\frac{dw_i}{d\chi^2}G(\chi,\theta),
\end{equation}
and we invoked the identity
$\left(\bm{n}\times\bm{v}_{2}\right)^2=\theta^2$. The minus sign in
Eq.~(\ref{av-G}), i.e., the negativity of interference between the
currents, retraces to the general property of saturation of
radiation. In Eq.~(\ref{expand-sum-sq-J}), though, factor
(\ref{interf-JJ}) is multiplied by an $\omega$-dependent factor
$\cos \Phi$, and the corresponding product is yet to be integrated
over polar radiation angles. Given that $\cos \Phi$ is
sign-alternating, the spectrum may then actually oscillate with the
increase of $\Omega$, not only experience a suppression. Combining
with the previously computed bremsstrahlung contributions from
individual plates, which are $\omega$-independent, we get
\begin{eqnarray}\label{dIdomega-combined}
\left\langle\frac{dI}{d\omega}\right\rangle=\frac{2e^2}{\pi}\Bigg[\left\langle
F\right\rangle_1+\left\langle
F\right\rangle_2\qquad\qquad\qquad\qquad\qquad\nonumber\\
-4\int_0^{\infty}d\Theta^2\Theta^2\left\langle
G\right\rangle_1\left\langle
G\right\rangle_2\cos\Omega(1+\Theta^2)\Bigg].\,\,
\end{eqnarray}
The dependence of function (\ref{dIdomega-combined}) on $\Omega$ is
shown in Fig.~\ref{fig:Exact-oscill}. As anticipated, it exhibits an
oscillatory behavior.

To determine the magnitude and the phase of the oscillations, note
that at $\Omega\gg1$, the integral in the last term of
(\ref{dIdomega-combined}) is dominated by the vicinity of the lower
endpoint, and its evaluation with $\left\langle
G\right\rangle_{1,2}(\Theta)=\left\langle
G\right\rangle_{1,2}(0)+\mathcal{O}(\Theta^2)$ yields the
asymptotics
\begin{eqnarray}\label{dI-largelambda}
\left\langle\frac{dI}{d\omega}\right\rangle\underset{\Omega\to\infty}\simeq\frac{2e^2}{\pi}\Bigg[\left\langle
F\right\rangle_1+\left\langle
F\right\rangle_2\qquad\qquad\qquad\qquad\nonumber\\
+\frac{4\cos\Omega}{\Omega^2}\left\langle
G\right\rangle_1(0)\left\langle
G\right\rangle_2(0)+\mathcal{O}(\Omega^{-3})\Bigg].\,\,
\end{eqnarray}
According to this result, maxima of ${dI}/{d\omega}$ should appear
at
\begin{equation}\label{secondary-max}
\Omega\approx 2\pi n,
\end{equation}
with $n$ integer, independently of shapes of distribution functions
$\left\langle G\right\rangle$. That simple prediction is confirmed
by Fig.~\ref{fig:Exact-oscill}. A suggestive geometrical
interpretation for the rule (\ref{secondary-max}), in the spirit of
wave optics, is that the maxima correspond to situations when the
distance between the plates equals an integer number of full
coherence lengths $2\pi l_{\text{f}}$.

Concerning the main maximum (which accounts for the bulk of the
anti-LPM enhancement), however, there is no possibility to associate
it with the case of unit wavelength shift as proposed in
\cite{Blankenbecler}. For $\Omega\lesssim 1$, of course, formula
(\ref{dI-largelambda}) does not apply, but from the point of exact
formula (\ref{dIdomega-combined}), the main maximum of $dI/d\omega$
emerges when in the integrand of the last term, the maximum
of\footnote{The existence of a maximum for this expression at
non-zero $\Theta$ owes in particular to the factor $\Theta^2$, which
is due to the cancellation of interference effects at small $\Theta$
-- see the end of Sec.~\ref{sec:2scat}.} $\Theta^2\left\langle
G\right\rangle_1(\Theta)\left\langle G\right\rangle_2(\Theta)$,
achieved at some $\Theta=\Theta_{\star}(\Sigma)$, coincides with the
extremum of $\cos\Omega(1+\Theta^2)$:
\begin{equation}\label{denom=1+Theta^2}
\Omega_{\star}=\frac{\pi}{1+\Theta^2_{\star}},
\end{equation}
whereat $\cos\Theta_{\star}<0$. This situation has to be
discriminated from the case $\Omega=\pi$ in
Eq.~(\ref{dI-largelambda}), where the existence of a maximum for
$\Theta^2\left\langle G\right\rangle_1\left\langle G\right\rangle_2$
is irrelevant because there are several oscillations within its
width, and the region of dominant contribution shifts to the
endpoint. In the latter case, the half-wavelength shift corresponds
to a spectral \emph{minimum} next to the main maximum. As opposed to
that, case (\ref{denom=1+Theta^2}) with the account of the
denominator $1+\Theta^2_{\star}$ can be called the inclined
half-length, which is numerically close to quarter-wavelength.

It is worth stressing in this regard that although there is a
denominator in relation (\ref{denom=1+Theta^2}) depending on
$\Sigma$, but it obeys a relation $\Theta_{\star}(\Sigma)\lesssim
1$. Clearly, even at large $\Sigma$, the contributing $\Theta$ in
Eq.~(\ref{dIdomega-combined}) are $\Theta\lesssim 1$, i.e.,
$\theta\lesssim\gamma^{-1}$ with respect to the direction
$\bm{v}_2$. Therefore, the use of the in-medium coherence length
\cite{Galitsky-Gurevich}
\begin{equation}\label{l-f-in-medium}
l_{\text{c}}(\omega,\Sigma)=\frac{l_{\text{f}}(\omega)}{1+\Sigma^2}
\end{equation}
in the resonance condition $t_{21}\propto
l_{\text{c}}(\omega,\Sigma)$ is \emph{unjustified}. In fact,
employment of (\ref{l-f-in-medium}) in application to the present
problem led in \cite{BK-structured} to a large underestimate of the
location of the first spectral maximum.

Finally, one must pay attention to the pronounced spectral minimum
at $\omega=0$, which must be associated with zero wavelength shift.
This limit is known to correspond to a situation when the radiation
amplitude completely factorizes from the scattering one, and so
radiation on two plates becomes the same as on one plate with the
aggregate thickness. Such a relation may be not immediately
transparent from Eq.~(\ref{dIdomega-combined}), although proves by
straightforward integration; we will also supply a simpler
derivation thereof in Sec.~\ref{subsec:V0}.

\subsection{Weighting with Gaussian distribution}

To deduce all the consequences from the averaging, let us now assume
$\frac{dw}{d^2\chi}$ in (\ref{av-G}) to be a Gaussian weighting
distribution (\ref{Gauss-distr}). The averaging of $G$ with such a
distribution simplifies after an integration by parts:
\begin{subequations}
\begin{eqnarray}
\left\langle G \right\rangle_{\sigma}(\Theta)&=&\int_0^\infty d\chi^2 \frac1{\sigma^2}e^{-\chi^2/\sigma^2}G\nonumber\\
&\equiv& \int_0^\infty d\chi^2 e^{-\chi^2/\sigma^2}\!\frac{\partial}{\partial\chi^2}G\label{J-scalar-av-chi-a}\\
&=&\int_0^\infty dX
\frac{e^{-{X}/{\Sigma^{2}}}}{\left[(X+1-\Theta^2)^2+4\Theta^2\right]^{3/2}}.\qquad\label{J-scalar-av-chi}
\end{eqnarray}
\end{subequations}
The behavior of function (\ref{J-scalar-av-chi}) is illustrated in
Fig.~\ref{fig:J}(b) for several values of $\Sigma$. It is largely
similar to that of $G$ shown in Fig.~\ref{fig:J}(a), but is smoother
at $\Theta\sim\Sigma\gg 1$, the latter property being natural
inasmuch as the unit step function in (\ref{J-largeangle-approx}) is
smoothened out by the averaging over $\chi$.

Numerically, Eq.~(\ref{dIdomega-combined}) with entries
(\ref{Fsigma-change-var}), (\ref{J-scalar-av-chi}) proves to be
equivalent to Eqs.~(2.34), (2.36) of \cite{BK-structured} under the
correspondence rule (\ref{b-Sigma2}), although analytically, the
equivalence of both approaches is challenging to verify. Our
representation seems to be better suited for the analytic study of
$\Sigma$- and $\Omega$-dependences. With its aid, we will now derive
small- and large-$\Sigma$ asymptotics of function $\left\langle G
\right\rangle_{\sigma}$, which will be useful in what follows.

At small $\Sigma$, plugging expansion (\ref{J-quadrupole-approx-a})
to Eq.~(\ref{J-scalar-av-chi-a}), and integrating termwise, we
obtain
\begin{subequations}
\begin{eqnarray}
\left\langle
G\right\rangle_{\sigma}(\Theta)&=& \frac{1}{4\Theta^2}\sum_{n=0}^{\infty}\frac{(n+1)!\Sigma^{2(n+1)}}{(\Theta^2+1)^{n+1}}\Bigg[P_n\left(\frac{\Theta^2-1}{\Theta^2+1}\right)\nonumber\\
&\,&\qquad\quad-\frac{\Theta^2-1}{\Theta^2+1}P_{n+1}\left(\frac{\Theta^2-1}{\Theta^2+1}\right)\Bigg]\label{G-small-Sigma-series}\\
&\underset{\Sigma\ll
1}{\simeq}&\Sigma^2(1+\Theta^2)^{-3}+\mathcal{O}(\Sigma^4).\label{G-small-Sigma}
\end{eqnarray}
\end{subequations}
Similarly to Eq.~(\ref{F-aver-series}), expansion
(\ref{G-small-Sigma-series}) diverges, but again, it remains
sensible as an asymptotic series. Its leading term
(\ref{G-small-Sigma}) coincides with the asymptotics of non-averaged
expression (\ref{J-quadrupole-approx}), granted that there the
$\chi$-dependence factorizes. The behavior of approximation
(\ref{G-small-Sigma}) is shown in Fig.~\ref{fig:J}(b) by dashed
curves.

On the other hand, at significant $\Sigma$, it is legitimate to
expand $e^{-{X}/{\Sigma^{2}}}$ to Taylor series in vicinity of point
$X=\Theta^2-1$ and integrate termwise:
\begin{eqnarray}\label{Gsigma-la}
\left\langle G\right\rangle_{\sigma}(\Theta) &\underset{\Sigma\gg
1}{\approx}& e^{-\frac{\Theta^2-1}{\Sigma^{2}}}\int_0^\infty dX
\frac{1}{\left[(X+1-\Theta^2)^2+4\Theta^2\right]^{3/2}}\nonumber\\
&\,&\qquad\qquad\qquad\qquad\times \left(1+\frac{\Theta^2-1-X}{\Sigma^2}\right)\nonumber\\
&=&
\frac{1}{1+\Theta^2}e^{-\frac{\Theta^2-1}{\Sigma^{2}}}\left(\frac12-\frac1{\Sigma^2}\right)\nonumber\\
&\approx&\frac{1}{2(1+\Theta^2)}e^{-\frac{1+\Theta^2}{\Sigma^{2}}}.
\end{eqnarray}
According to Fig.~\ref{fig:J}(b) (dot-dashed curves), approximation
(\ref{Gsigma-la}) works well for $\Sigma > 3$. At large $\Sigma$ and
fixed $\Theta$, the exponential here tends to unity, and form
(\ref{J-largeangle-approx}) is retrieved.


\subsection{Averaging with Moli\`{e}re distribution. Impact parameter representation}

If instead of a Gaussian we prefer to use Moli\`{e}re  weighting
distribution (\ref{Moliere-distr-def}), it may be more convenient to
return to Eq.~(\ref{expand-sum-sq-J}) for the radiation spectrum. In
the last line of Eq.~(\ref{expand-sum-sq-J}), the product of
currents  consists of 4 terms, which have similar structure. It
suffices to calculate only one of those,
\begin{eqnarray}\label{89}
O_{13}&=&\int d^2n \frac{\bm{n}-\bm{v}_1}{\gamma^{-2}+(\bm{n}-\bm{v}_1)^2}\cdot \frac{\bm{n}-\bm{v}_3}{\gamma^{-2}+(\bm{n}-\bm{v}_3)^2}\nonumber\\
&\,&\qquad\qquad\times\cos\frac{\omega t_{21}}2
\left[\gamma^{-2}+(\bm{n}-\bm{v}_2)^2\right],
\end{eqnarray}
while the other terms can be reconstructed by replacements
$\bm{v}_3\to \bm{v}_2$ or/and $\bm{v}_1\to \bm{v}_2$.

Changing the integration variable to
$\bm{n}'_{\perp}=\bm{n}-\bm{v}_2$, inserting representations
(\ref{wf-angle-wf-imp-par}) for the first factor in the first line
of (\ref{89}), and a complex conjugate representation for the second
factor, one brings it to the form of an integral over impact
parameters:
\begin{eqnarray}\label{O31-ReS}
O_{13}&=&
\int d^2 r_1 e^{i(\bm{v}_2-\bm{v}_1)\cdot \bm{r}_1} \frac{\partial}{\partial \bm{r}_1}K_0(r_1/\gamma)\nonumber\\
&\,&\cdot\int d^2 r_3 e^{i(\bm{v}_3-\bm{v}_2)\cdot \bm{r}_3} \frac{\partial}{\partial \bm{r}_3}K_0(r_3/\gamma)\nonumber\\
&\,&\qquad\qquad\qquad\times
\mathfrak{Re}S_{\omega}\left(\bm{r}_1-\bm{r}_3,t_{21}\right).
\end{eqnarray}
Here
\begin{equation}\label{}
S_{\omega}\!\left(\bm{r}_1-\bm{r}_3,t_{21}\right) 
=\frac1{(2\pi)^2}\!\int d^2n'_{\perp} e^{i\bm{n}'_{\perp}\cdot
(\bm{r}_1-\bm{r}_3)-i\frac{\omega t_{21}}2\!
\left(\gamma^{-2}+\bm{n}'^2_{\perp}\right)}
\end{equation}
may be regarded as Green's function for a two-dimensional free
Schr\"{o}dinger equation on the light front:
\begin{eqnarray}\label{}
\left[\frac{\omega}2\!\left(\!\gamma^{-2}\!-\!\frac{\partial^2}{\partial(\bm{r}_1\!-\!\bm{r}_3)^2}\!\right)-i\frac{\partial}{\partial
t_{21}}\right]\!S_{\omega}\!\left(\bm{r}_1-\bm{r}_3,t_{21}\right)\!\Theta(t_{21})\nonumber\\
=-i\delta(\bm{r}_1-\bm{r}_3)\delta(t_{21}),\qquad
\end{eqnarray}
and it evaluates in closed form:
\begin{equation}\label{}
\mathfrak{Re} S_{\omega}\left(\bm{r}_1-\bm{r}_3,t_{21}\right)
=\frac1{2\pi\omega
t_{21}}\sin\left[\frac{(\bm{r}_1-\bm{r}_3)^2}{2\omega
t_{21}}-\frac{\omega t_{21}}{2\gamma^{2}}\right].
\end{equation}
Note that this function depends only on the difference between
$\bm{r}_1$ and $\bm{r}_3$, just as in (\ref{dIdomega-combined}) the
cosine depends, besides $\omega t_{21}$, only on the angle between
$\bm{v}_2$ and $\bm{n}$.

Restoring the rest of the interference terms by substitutions
$\bm{v}_1\to \bm{v}_2$ or $\bm{v}_3\to \bm{v}_2$, and combining
them, we get
\begin{eqnarray}\label{}
\nonumber\\
\frac{dI}{d\omega}=\frac{dI_1}{d\omega}+\frac{dI_2}{d\omega}\qquad\qquad\qquad\qquad\qquad\qquad\qquad\qquad \nonumber\\
+\frac{2e^2}{\pi^2}\int d^2 r_1 d^2r_3
\left(1-e^{i(\bm{v}_1-\bm{v}_2)\cdot \bm{r}_1}
\right)\left(1-e^{i(\bm{v}_2-\bm{v_3})\cdot \bm{r}_3}
\right) \nonumber\\
\times \frac{\partial}{\partial \bm{r}_1}K_0(r_1/\gamma)
\cdot\frac{\partial}{\partial
\bm{r}_3}K_0(r_3/\gamma)\mathfrak{Re}S_{\omega}\left(\bm{r}_1-\bm{r}_3,t_{21}\right).
\end{eqnarray}
Convolving this with the Moli\`{e}re distribution function, and
integrating over azimuths of $\bm{r}_1$ and $\bm{r}_3$, leads to the
result
\begin{eqnarray}\label{impact-2scat}
\left\langle\frac{dI}{d\omega}\right\rangle_{\text{M}}=\left\langle\frac{dI_1}{d\omega}\right\rangle_{\text{M}}+\left\langle\frac{dI_2}{d\omega}\right\rangle_{\text{M}} \qquad\qquad\qquad\qquad\quad\nonumber\\
+\frac{2e^2}{\pi\Omega}
\int_0^{\infty} d \rho_1\rho_1 K_1(\rho_1) \left\{1-e^{-n_1l_1 \int d\sigma(\chi)[1-J_0(\rho_1\gamma \chi)]}\right\} \nonumber\\
\times\int_0^{\infty} d \rho_3\rho_3 K_1(\rho_3) \left\{1-e^{-n_2l_2
\int d\sigma(\chi)[1-J_0(\rho_3\gamma\chi)]}\right\}
\nonumber\\
\times
\cos\left(\frac{\rho_1^2+\rho_3^2}{4\Omega}-\Omega\right)J_1\left(\frac{\rho_1\rho_3}{2\Omega}\right),\quad
\end{eqnarray}
where for convenience we rescaled the impact parameter variable to
$\rho_{1,3} =r_{1,3}/\gamma$. Note that after integration over the
azimuth, the dependence only on the impact parameter difference is
somewhat disguised.

At large $\Omega$, upon approximations
$\cos\left(\frac{\rho_1^2+\rho_3^2}{4\Omega}-\Omega\right)\to\cos\Omega$,
$J_1\left(\frac{\rho_1\rho_3}{2\Omega}\right)\to
\frac{\rho_1\rho_3}{4\Omega}$, (\ref{impact-2scat}) factorizes:
\begin{eqnarray}\label{}
\left\langle\frac{dI}{d\omega}\right\rangle_{\text{M}}\underset{\Omega\to\infty}\to\left\langle\frac{dI_1}{d\omega}\right\rangle_{\text{M}}+\left\langle\frac{dI_2}{d\omega}\right\rangle_{\text{M}}\qquad\qquad\qquad\qquad\quad\nonumber\\
+\frac{e^2\cos\Omega}{2\pi\Omega^2}
\int_0^{\infty}\! d \rho_1\rho_1^2 K_1(\rho_1)\! \left\{1-e^{-n_1l_1 \int d\sigma(\chi)[1-J_0(\rho_1\gamma \chi)]}\right\} \nonumber\\
\times\int_0^{\infty} d \rho_3\rho_3^2 K_1(\rho_3)
\left\{1-e^{-n_2l_2 \int
d\sigma(\chi)[1-J_0(\rho_3\gamma\chi)]}\right\}.\,\,
\end{eqnarray}
This form proves to be equivalent to Eq.~(\ref{dI-largelambda}), by
virtue of the identity
\begin{subequations}
\begin{eqnarray}
\frac14\int_0^{\infty} d \rho\rho^2 K_1(\rho) \left\{1-e^{-nl \int d\sigma(\chi)[1-J_0(\rho\gamma\chi)]}\right\}\quad\label{80a}\\
=\frac14\int d^2\chi \frac{dw_{\text{M}}}{d^2\chi} \int_0^{\infty} d \rho\rho^2 K_1(\rho) \left[1-J_0(\rho\gamma\chi)\right]\nonumber\\
=\int d^2\chi \frac{dw_{\text{M}}}{d^2\chi}
\left[\frac12-\frac1{2(1+X)^2}\right] \equiv\left\langle
G\right\rangle_{\text{M}}(0)\quad
\end{eqnarray}
\end{subequations}
[cf. Eq.~(\ref{G(0)})].

\section{Analysis of the interference pattern}\label{sec:analysis}

The spectrum of radiation on two plates that we have evaluated, even
under the simplification of a negligible plate thicknesses, is a
function of 3 variables: $\Sigma_1$, $\Sigma_1$, and $\Omega$. In
limiting cases, though, it reduces to simpler functions, which will
be scrutinized below.

\begin{figure}
\includegraphics{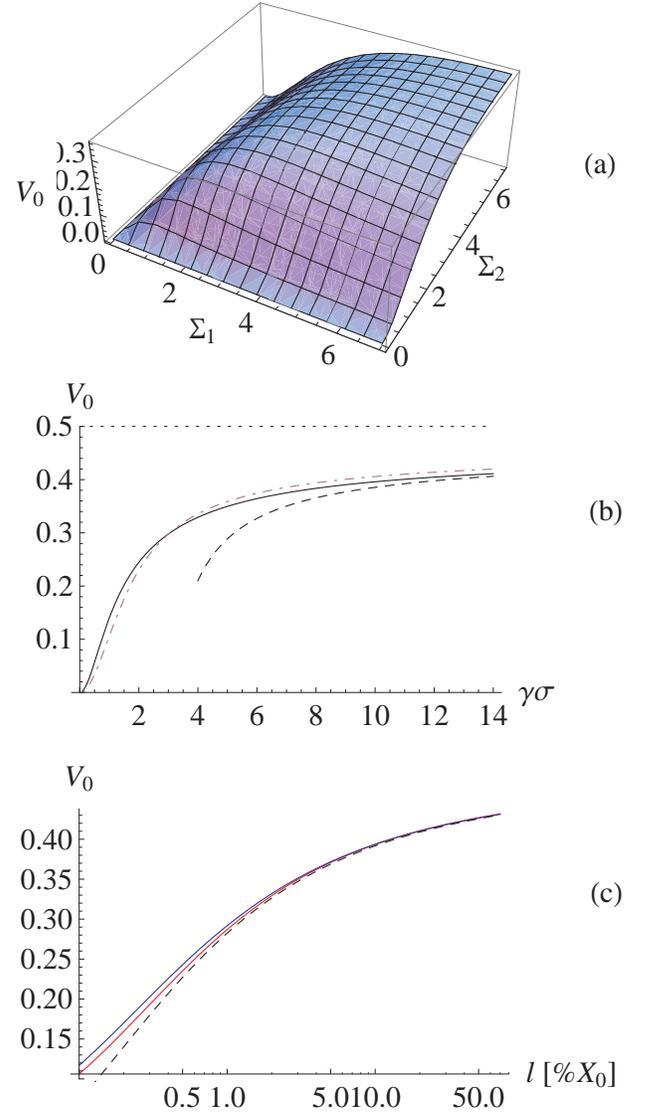}
 \caption{\label{fig:V} The visibility of the low-$\omega$ dip, Eq.~(\ref{V-through-F}). (a) For Gaussian averaging and arbitrary $\Sigma_1$, $\Sigma_2$. (b) For Gaussian averaging and $\Sigma_1=\Sigma_2=\Sigma$. Solid black curve, exact expression~(\ref{visibility}). Dashed black curve, its asymptotics (\ref{V-largeSigma}). Purple dot-dashed curve, the visibility in the Blankenbecler-Drell approximation. (c) For Moli\`{e}re averaging. Blue curve, carbon; red curve, gold; black dashed curve, Gaussian averaging with rms scattering angle computed by formula (\ref{Highland-approx}).}
\end{figure}

\subsection{Visibility of the low-$\omega$ dip}\label{subsec:V0}

For experimental tests, it is imperative to estimate visibilities of
gross features of the spectrum, and find their optimum by choosing
appropriate foil materials and their thicknesses. The interference
term reaches its largest absolute value at $\omega=0$. Therefore, it
is natural to consider the characteristic quantity
\begin{equation}\label{V-def}
V_0\left(\Sigma_1,\Sigma_2\right)=1-\left\langle\frac{dI}{d\omega}\right\rangle\bigg|_{\omega=0}\bigg/\left\langle\frac{dI}{d\omega}\right\rangle\bigg|_{\omega=\infty}.
\end{equation}
Adopting the terminology from optics, this may be called the
visibility of the low-$\omega$ dip. With the aid of
Eq.~(\ref{dIdomega-combined}), it can be explicitly written as
\begin{subequations}\label{visibility}
\begin{equation}\label{visibility0}
V_0=\frac{4\int_0^{\infty}d\Theta^2\Theta^2\left\langle
G\right\rangle_1(\Theta)\left\langle
G\right\rangle_2(\Theta)}{\left\langle F\right\rangle_1+\left\langle
F\right\rangle_2}.
\end{equation}
As we already mentioned at the end of Sec.~\ref{subsec:Aggr-spectr},
here the spectrum must reduce to that from a single plate of
aggregate thickness. This can now be proven rather easily, based on
the impact parameter representation (\ref{impact-2scat}). There,
$S_{\omega}\underset{\Omega\to0}\to\delta(\bm{r}_1-\bm{r}_3)$, and
terms $\frac{dI_1}{d\omega}$, $\frac{dI_2}{d\omega}$ cancel with
some of the interference terms, leaving $\frac{dI}{d\omega}
\underset{\Omega\to 0}\to \frac{dI_{l_1+l_2}}{d\omega}$. Thereby,
$V_0$ indeed reduces to the form
\begin{equation}\label{V-through-F}
V_0=1-\frac{\left\langle F\right\rangle_{l_1+l_2}}{\left\langle
F\right\rangle_1+\left\langle F\right\rangle_2}.
\end{equation}
\end{subequations}

At small $\Sigma$,
\begin{equation}
V_0\underset{\Sigma_{1,2}\ll
1}\simeq\frac35\frac{1}{\Sigma^{-2}_1+\Sigma^{-2}_2},
\end{equation}
i.e., essentially it is proportional to the minimal among
$\Sigma^{2}_1$, $\Sigma^{2}_2$.

At large $\Sigma_1$, $\Sigma_2$,
\begin{equation}\label{V-largeSigma}
V_0\underset{\Sigma_{1,2}\gg
1}\simeq\frac12-\frac{\ln\left(\Sigma_1/\Sigma_2+\Sigma_2/\Sigma_1\right)
}{2(\ln\Sigma_1\Sigma_2-\gamma_{\text{E}}-1)}.
\end{equation}
Thus, in principle, asymptotically $V_0$ tends to $\frac12$. That is
natural, inasmuch as radiation at one plate saturates as a function
of its thickness, and hence the radiation at any two strongly
scattering foils must be about twice stronger than on one of them.
But practically, such a saturation is achieved only logarithmically,
and is too remote, so the second term in Eq.~(\ref{V-largeSigma}) is
usually significant.

The behavior of (\ref{visibility}) and asymptotic approximation
(\ref{V-largeSigma}) is illustrated in Fig.~\ref{fig:V}(b) by black
solid and dashed curves. For comparison, the purple dot-dashed curve
also shows the corresponding result for Blankenbecler's theory, when
in Eq.~(\ref{V-through-F}) $\left\langle F\right\rangle_{\sigma}$ is
replaced by $F(\Sigma/2)$. Notably, that curve intersects with the
exact result at $\Sigma\approx 3$, and keeps close to it at greater
$\Sigma$. But at $\Sigma\sim 1$ the difference is relatively large.
It is also interesting to note that the Blankenbecler-Drell
prediction for $V_0$ is lower than the exact one, i.e., the neglect
of fluctuations leads to an underestimation of the interference
effect in radiation. That means that under the conditions of
cancellation of dipole contributions in the interference,
fluctuations help creating uncompensated contributions.

Likewise, it is useful to compare in this limit the predictions of
Gaussian and Moli\`{e}re averaging [see Fig.~\ref{fig:V}(c)]. It
appears that the Gaussian averaging prediction becomes inaccurate at
$l\lesssim 10^{-3}X_0$, just where the accuracy of approximation
(\ref{Highland-approx}) itself becomes worse than 10\%.

\subsection{Visibility of secondary minima and maxima}

\begin{figure}
\includegraphics{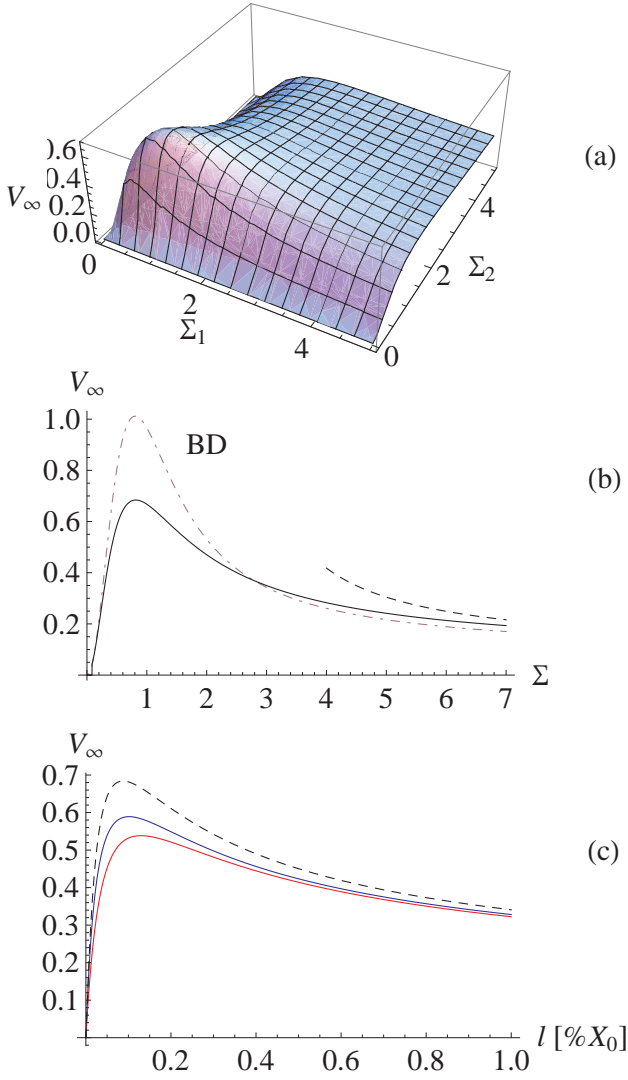}
 \caption{\label{fig:Vinfty} Same as Fig.~\ref{fig:V}, for visibility of secondary minima and maxima, Eq.~(\ref{V_infty-def}). The dashed curve in (b) obeys Eq.~(\ref{Vinfty-asympt}) for $\Sigma_1=\Sigma_2=\Sigma$.}
\end{figure}

In the opposite limit of large $\omega$, according to
Eq.~(\ref{dI-largelambda}), the visibility may be characterized by
the ratio
\begin{equation}\label{V_infty-def}
V_{\infty}\left(\Sigma_1,\Sigma_2\right)=\frac{4\left\langle
G\right\rangle_1(0)\left\langle G\right\rangle_2(0)}{\left\langle
F\right\rangle_1+\left\langle F\right\rangle_2}.
\end{equation}
At small $\Sigma$, this quantity behaves as
\begin{equation}\label{}
V_{\infty}\simeq\frac{12}{\Sigma_1^{-2}+\Sigma_2^{-2}},
\end{equation}
whereas at large $\Sigma_1$, $\Sigma_2$, it logarithmically
decreases:
\begin{equation}\label{Vinfty-asympt}
V_{\infty}\simeq\frac{1}{2(\ln\Sigma_1\Sigma_2-\gamma_{\text{E}}-1)}.
\end{equation}
The dependence of $V_{\infty}$ on both parameters $\Sigma_1$ and
$\Sigma_2$ is shown in Fig.~\ref{fig:Vinfty}(a). Its maximum is
achieved at $\Sigma_1=\Sigma_2\approx1$ [see
Fig.~\ref{fig:Vinfty}(b)]. There, the function still does not exceed
0.7. Yet, in (\ref{dI-largelambda}) it is multiplied by
$\Omega^{-2}$ with $\Omega\gtrsim\pi$, hence, actual visibility in
the high-$\omega$ region is on the level of a few percent, demanding
formidable measurement statistics.

For comparison, in Fig.~\ref{fig:Vinfty}(b) by dot-dashed curve we
show the behavior of Eq.~(\ref{V_infty-def}) after replacement of
averaged radiation formfactors by non-averaged functions of averaged
argument:
\begin{eqnarray}\label{89}
\left\langle F\right\rangle_{\sigma}\to F(\Sigma/2), \quad
\left\langle G\right\rangle_{\sigma}(0)\to
G(\Sigma,0)=\frac12-\frac{1}{2(1+\Sigma^2)^2}\\ \text{(LP-BD)},
\qquad\qquad\qquad\qquad\qquad \nonumber
\end{eqnarray}
which we presume to correspond to Blankenbecler-Drell approximation.
Function (\ref{89}) intersects with the exact result at
$\Sigma\approx2.5$, and keeps close to it at greater $\Sigma$, but
at $\Sigma\approx1$ the difference is rather large.

Fig.~\ref{fig:Vinfty}(c) compares predictions of
Eq.~(\ref{V_infty-def}) when averaging is performed with the
Moli\`{e}re distribution, by Eqs.~(\ref{80a}), (\ref{exp-K1}) (the
blue solid curve for carbon, the red solid curve for gold), and with
Gaussian distribution, Eqs.~(\ref{J-scalar-av-chi}) and
(\ref{Fsigma-change-var}) (black dashed curve). There is a distinct
difference between the predictions around the maximum, which is
achieved at $l\approx 10^{-3}X_0$, i.e., $\Sigma\approx 1$. The sign
of this difference tells that in the region of substantial photon
energies, the account of fluctuations of scattering leads to a
lowering of the interference effect in radiation, contrary to the
situation with $V_0$ in the previous section. That admits rather
simple explanation: fluctuations always tend to suppress the
radiation, but at $\omega=0$, where the interference effect itself
is suppressive, that worked towards increasing the effect, whereas
in the large-$\omega$ region it just lowers both amplitudes
$\left\langle G\right\rangle_1(0)$ and $\left\langle
G\right\rangle_2(0)$ in the interference term.

\subsection{Spectrum oscillation shapes}

For a generic case, integral (\ref{dIdomega-combined}) can only be
evaluated numerically. But for limiting cases
(\ref{J-quadrupole-approx}) and (\ref{J-largeangle-approx}), the
integration is analytically manageable. For completeness, in the
remainder of this section, we will collect formulae for the
corresponding limiting spectral shapes, and compare them with
results of \cite{BK-structured}, obtained for same limits, at
$\sigma_1=\sigma_2$.

\subsubsection{Small-angle scattering in both plates}

\begin{figure}
\includegraphics{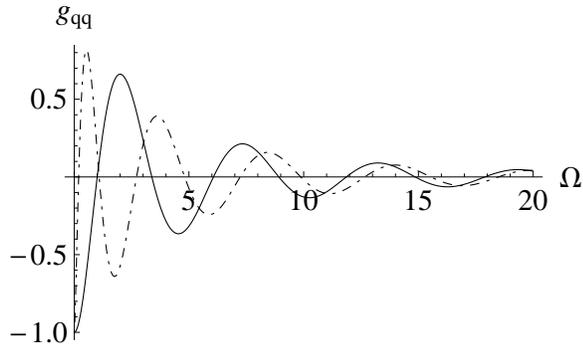}
 \caption{\label{fig:g-qq} Interference function for the case of double small-angle-scattering. Solid curve, Eq.~(\ref{g-qq-def}) -- close to the red curve of Fig.~(\ref{fig:Exact-oscill}). Blue dot-dashed curve, Baier-Katkov's function $\frac{10}{3}G(\Omega)$ \cite{BK-structured}.}
\end{figure}

To begin with, consider the case when scattering angles in both of
the foils are small. Inserting (\ref{G-small-Sigma}) to
(\ref{dIdomega-combined}), we get
\begin{eqnarray}\label{qq}
\left\langle\frac{dI}{d\omega}\right\rangle&\underset{\Sigma_{1,2}\ll1}\simeq&\left(\left\langle\frac{dI_1}{d\omega}\right\rangle+\left\langle\frac{dI_2}{d\omega}\right\rangle\right)\nonumber\\
&\,&\times\left(1+\frac35\frac{1}{\Sigma_1^{-2}+\Sigma_2^{-2}}g_{\text{qq}}(\Omega)\right),
\end{eqnarray}
where
\begin{equation}\label{}
\left\langle\frac{dI_1}{d\omega}\right\rangle\simeq\frac{2e^2}{3\pi}\Sigma_1^2\left(1-\frac3{10}\Sigma_1^2\right),
\end{equation}
and
\begin{eqnarray}\label{g-qq-def}
g_{\text{qq}}(\Omega)=-20\int_0^{\infty}\frac{d\Theta^2\Theta^2}{(1+\Theta^2)^6}\cos\Omega(1+\Theta^2),\\
\qquad g_{\text{qq}}(0)=-1\qquad\qquad\qquad\nonumber
\end{eqnarray}
is the quadrupole-quadrupole interference function. The latter
function achieves its first (anti-LPM) maximum at $\Omega\approx 2$
(see Fig.~\ref{fig:g-qq}). According to Eqs.~(\ref{qq}),
(\ref{g-qq-def}), the visibility of spectral fringes is
$\sim\frac3{10}\Sigma_1^2$.

It must be noted that our function (\ref{g-qq-def}) appears to
differ from function $\frac{10}3 G(T)$ obtained in
\cite{BK-structured}, Eq.~(2.37). The comparison of those functions
in Fig.~\ref{fig:g-qq} shows ample differences except in the limits
$\Omega=0$ and $\Omega\to\infty$. We believe therefore that the
corresponding result of \cite{BK-structured} is in
error.\footnote{In order to derive the corresponding result from
equations (2.34), (2.36) of \cite{BK-structured}, one has to expand
them up to $b^{-2}$ (while contributions $\sim b^{-1}$ from
$2{dw^{(2)}_{br1}}/{d\omega}$ and ${dw^{(2)}_{br3}}/{d\omega}$ must
cancel). We did not follow this procedure, wherefore we can not
indicate the source of the error.}


\subsubsection{Radiation at double large-angle scattering}

\begin{figure}
\includegraphics{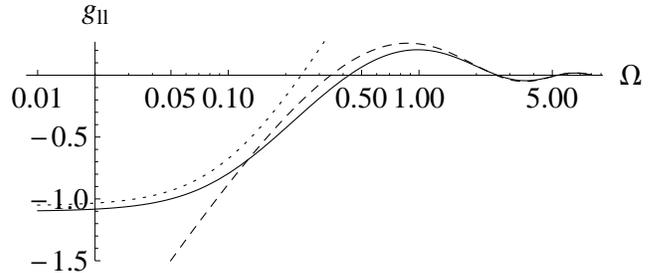}
 \caption{\label{fig:gll} Interference function for the case of double large-angle-scattering.
 Solid curve, Eq.~(\ref{E1}) -- close to the blue curve of Fig.~\ref{fig:Exact-oscill}.
 Dashed curve, $g_{\text{ll}}(0,\Omega)$. Dotted curve, asymptotics~(\ref{g_ll-approx}).}
\end{figure}

The next case to be considered is when scattering angles on each of
the foils are $\gg\gamma^{-1}$. Inserting Eq.~(\ref{Gsigma-la}) to
(\ref{dIdomega-combined}), obtains
\begin{eqnarray}\label{ll}
\left\langle\frac{dI}{d\omega}\right\rangle\underset{\Sigma_{1,2}\gg1}\simeq\frac{2e^2}{\pi}\Big[
\ln \Sigma^2_1+\ln
\Sigma^2_2-2\gamma_{\text{E}}-2\qquad\quad \nonumber\\
+g_{\text{ll}}\left(\Sigma_1^{-2}+\Sigma_2^{-2},\Omega\right)\Big],
\end{eqnarray}
where
\begin{eqnarray}\label{E1}
g_{\text{ll}}(\xi,\Omega)&=&-\mathfrak{Re}\left[\int_0^{\infty}\frac{d\Theta^2\Theta^2}{(1+\Theta^2)^2}e^{-(1+\Theta^2)(\xi+i\Omega)}\right]\nonumber\\
&=&e^{-\xi}\cos\Omega-\mathfrak{Re}\left[(1+\xi+i\Omega)E_1\left(\xi+i\Omega\right)\right],\nonumber\\
\end{eqnarray}
with $E_1(z)=\int_1^{\infty}dt\frac{e^{-zt}}{t}$ the exponential
integral. Term $\xi=\Sigma_1^{-2}+\Sigma_2^{-2}$ in the argument of
$E_1$ is negligible if $\Omega\gg \Sigma_1^{-2}+\Sigma_2^{-2}\ll 1$;
then our Eq.~(\ref{ll}) reduces to Eq.~(2.45) of
\cite{BK-structured}. This limiting function has an anti-LPM maximum
at $\Omega\approx 0.9$ (see Fig.~\ref{fig:gll}), which agrees with
the conclusions of Blankenbecler-Drell and Baier-Katkov that the
maximum is achieved at $\Omega\approx 1$.\footnote{However, we
disagree with their interpretation of it as a unit-wavelength
resonance. As was argued above, it is rather an inclined
half-wavelength. Besides that, Baier and Katkov conducted their
evaluation of the maximum for a specific value of the target
thickness 11 $\mu$m.} However, for small $\Omega$, term $\xi$ in the
argument of $E_1$ is crucial. There, $E_1(z)\simeq-\ln
z-\gamma_{\text{E}}$, wherewith
\begin{equation}\label{g_ll-approx}
g_{\text{ll}}(\xi,\Omega)\underset{\xi\ll
1}\approx\frac{1+\xi}2\ln\left(\xi^2+\Omega^2\right)+\gamma_{\text{E}}+1-\xi.
\end{equation}
Corrections proportional to $\xi$, especially in the prefactor of
the logarithm in (\ref{g_ll-approx}), can be important at moderately
small $\Sigma_1^{-2}+\Sigma_2^{-2}$. Neglecting $\Omega^2$ or
$\xi^2$ under the logarithm sign in (\ref{g_ll-approx}), we arrive
correspondingly at Eq.~(2.42) or (2.43) of \cite{BK-structured}. In
Fig.~\ref{fig:gll} we compare the behavior of function (\ref{E1})
(solid curve), with that of approximation (\ref{g_ll-approx})
(dotted curve), and of an approximation resulting when we put in
Eq.~(\ref{E1}) $\Theta\to 0$ (dashed curve). The latter two
approximations are seen to work in complementary regions.

\subsubsection{Asymmetric case}

Finally, we examine the case when one of the foils scatters weakly,
whereas another one scatters the electrons through angles
$\gg\gamma^{-1}$. Inserting the corresponding limiting forms
(\ref{G-small-Sigma}) and (\ref{Gsigma-la}) to
(\ref{dIdomega-combined}), we arrive at
\begin{equation}\label{}
\left\langle\frac{dI}{d\omega}\right\rangle\underset{\Sigma_2\gg1}{\underset{\Sigma_1\ll1}\simeq}\frac{2e^2}{\pi}\left\{
\ln
\Sigma^2_2-\gamma_{\text{E}}-1+\frac{\Sigma^2_1}3\left[1+g_{\text{ql}}(\Omega)\right]\right\},
\end{equation}
where
\begin{eqnarray}\label{g_ql-def}
g_{\text{ql}}(\Omega)=-6\int_0^{\infty}\frac{d\Theta^2\Theta^2}{(1+\Theta^2)^4}\cos\Omega(1+\Theta^2),\\
g_{\text{ql}}(0)=-1.\qquad \qquad \nonumber
\end{eqnarray}
This function has a maximum at $\Omega\approx 1.6$.

Notably, there is no logarithmic dependence in the interference
term, and its amplitude does not depend on $\Sigma_2$, which
corresponds to strict saturation of radiation. Correspondingly, the
visibility in this case is $\sim\frac{\Sigma^2_1}{\ln \Sigma_2^2}$,
i.e., even lower than for the case of double small-angle scattering.

\section{Account for foil thickness(es)}\label{sec:plate-formfactor}

Hitherto we presumed negligible thicknesses of both targets, but at
a sufficiently large $\omega$, finite target thickness must
inevitably show up. Indeed, when $l_{f}$ decreases down to the value
of $l$, the target geometry gets resolved. Thereat, the radiation
spectrum must additionally rise from the doubled but logarithmically
saturated values to the completely saturation-free Bethe-Heitler
value.

In principle, our framework requires only minor modification  to
reflect the mentioned rise: It suffices to multiply the radiation
spectrum computed above by the plate formfactor, which smoothly
interpolates between unity at $\omega\lesssim
\frac{2\gamma^2}{t_{21}}$ and Migdal's function $\Phi_M$ at
$\omega\gg\frac{2\gamma^2}{t_{21}}$ (specifically, $\omega\gtrsim
e^2\frac{\gamma^2X_0}{l^2}$). The implementation of such an
interpolation, however, is beyond our scope in the present paper.

\section{Comparison with experiment}\label{sec:experiment}

The best way to test our equations is to confront them with
experimental data. Measurements of bremsstrahlung spectra from 178
GeV electrons on a sequence of plates were performed in
\cite{Andersen-2foils,NA63-plans} with two equal 26 $\mu$m thick
golden foils ($X_0^{\text{Au}}=3.4$ mm) separated by a gap of
variable width. The scattering strength parameter for one such a
foil, according to Eq.~(\ref{Highland-approx}), estimates as
$\gamma\sigma\approx2.7$, so the scattering angles there were rather
large.

\begin{figure}
\includegraphics{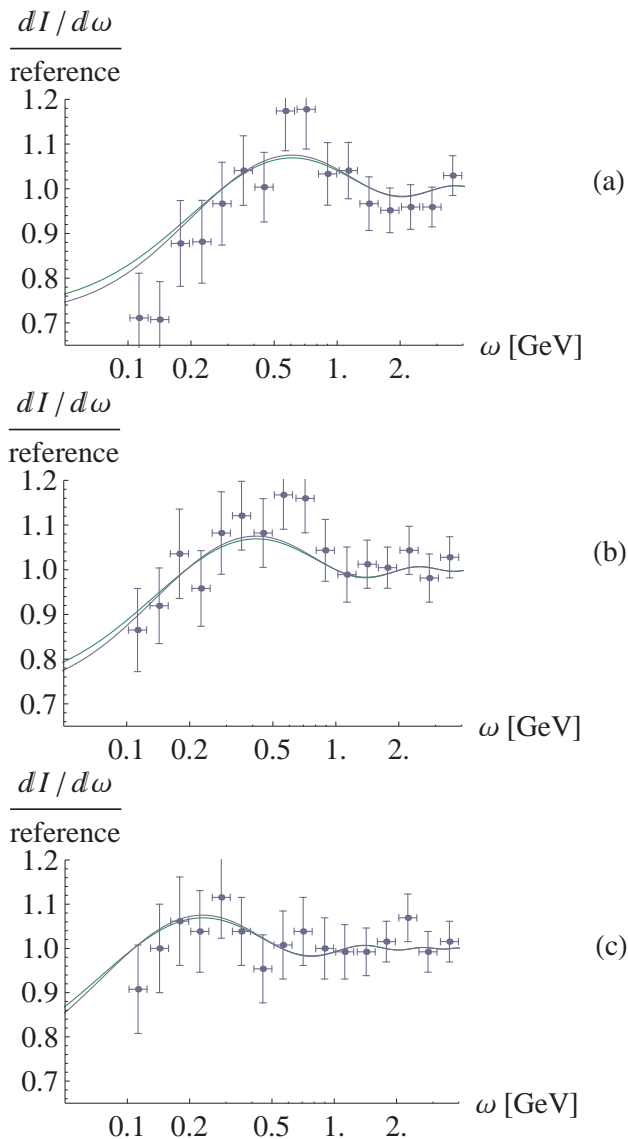}
 \caption{\label{fig:experiment} Shapes of spectral oscillations for conditions of experiment \cite{NA63-plans} ($l_1=l_2=26\mu$m). (a) $t_{21}=26+60\,\mu$m; (b) $t_{21}=26+100\,\mu$m; (c) $t_{21}=26+200\,\mu$m. Dark blue curves, Gaussian averaging, with $\Sigma$ calculated by Eq.~(\ref{Highland-approx}). Green curves, Moli\`{e}re averaging for gold.}
\end{figure}

In Fig.~\ref{fig:experiment} we compare predictions of our
Eq.~(\ref{dIdomega-combined}) and Eq.~(\ref{impact-2scat}) with the
experiment \cite{NA63-plans}, letting $t_{21}$ be equal to the
distance between plate centres, $l_{\text{g}}+l$. The agreement may
be regarded as fair. Some possible experimental inaccuracies at
$\omega\lesssim 50$ MeV were discussed in \cite{Andersen-2foils},
but they do not seem to cause large deviations.

Since our Gaussian and Moli\`{e}re averaging procedures give
numerically close results, and on the other hand, it was found in
Sec.~\ref{sec:1plate} that our prescriptions for scattering
parameters are close to those of Baier-Katkov and Zakharov, we
expect that Baier-Katkov's and Zakharov's predictions for the
radiation spectrum must be in agreement with the experiment
\cite{NA63-plans}, too, even though the qualitative inferences based
on the Baier-Katkov theory seem to disagree with it.

As for comparison of experimental results with predictions of
Blankenbecler, recalling our results of Sec.~\ref{sec:analysis}, it
should be noticed that the use of plates with scattering strength
$\gamma\sigma\approx 2.7$ closely corresponds to the point of
intersection of visibilities in BD approximation with exact ones
[see Fig.~\ref{fig:Vinfty}(b)]. Therefore, Blankenbecler's
predictions may agree with particular experiment \cite{NA63-plans}
rather nicely, too. Basic agreement of experimental results with
\cite{Blankenbecler} (and disagreement with
\cite{Blankenbecler-FunctionalInt}) was actually reported in
\cite{NA63-plans}. However, for foils several times thinner that
currently used (or made of a lighter material), we predict a growing
inaccuracy of Blankenbecler-Drell predictions. At the same time, for
weaker scattering foils, the visibility of \emph{secondary} spectral
minima and maxima can yet increase by about a factor of 2 [see
Figs.~\ref{fig:Vinfty}(b),(c)].



\section{Bremsstrahlung on randomly located plates}\label{sec:randomization}

\begin{figure}
\includegraphics{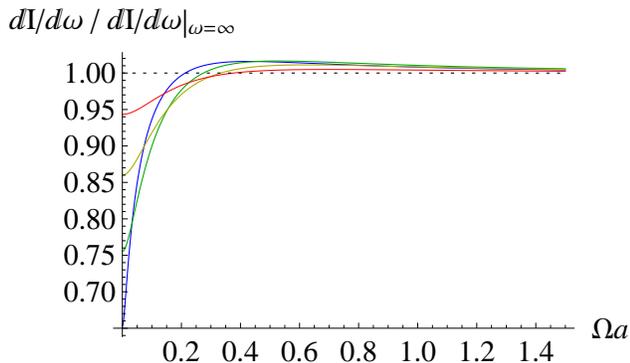}
 \caption{\label{fig:Oscill-aver-t21} Spectra of bremsstrahlung on two plates with a randomized distance between them, Eq.~(\ref{I-aver-t21}), for $\tau=3a$ and same parameter values for scattering strengths as in Fig.~\ref{fig:Exact-oscill}.}
\end{figure}

Having dwelt enough on the case of bremsstrahlung on two targets
separated by a fixed distance, it would be instructive also to
generalize it so that it could be compared with the conventional
case of bremsstrahlung in an extended random medium. Of course,
presently we possess equations for radiation only at two
scatterings, but arguably, the greatest contribution to the
interference must come from nearest scatterings. Therefore, we can
exploit previous formulae, if $t_{21}$ is regarded as a random
variable distributed within some range $\sim \tau$. Rather
realistically, it can be modeled by an exponential
$e^{-t_{21}/\tau}$, corresponding to the probability of avoiding an
encounter with the next atom in a uniform gas with the mean
interatomic distance $\tau$. To preserve the anti-LPM effect,
though, we must prevent $t_{21}$ from tending to zero, as was
pointed out in Sec.~\ref{sec:2scat}. For simplicity, we shall assume
the shape of the anticorrelation to be described also by an
exponential, but shorter-range one, thus choosing the distribution
function
\begin{equation}\label{dwdt21}
\frac{dw}{dt_{21}}=\frac1{\tau-a}\left(e^{-t_{21}/\tau}-e^{-t_{21}/a}\right).
\end{equation}
Parameter $a<\tau$ provides the `repulsion' distance between the
scatterers. Weighting Eq.~(\ref{dIdomega-combined}) with
(\ref{dwdt21}), we get
\begin{eqnarray}\label{I-aver-t21}
\left\langle\frac{dI}{d\omega}\right\rangle_{t_{21}}=\frac{2e^2}{\pi}\Bigg\{\left\langle
F\right\rangle_1+\left\langle
F\right\rangle_2\qquad\qquad\qquad\qquad\qquad\nonumber\\
+\frac4{\tau-a}\int_0^{\infty}d\Theta^2\Theta^2\left\langle
G\right\rangle_1\left\langle
G\right\rangle_2\qquad\qquad\qquad\qquad\qquad\nonumber\\
\times\left[-\frac{\tau}{1+(\frac{\omega\tau}{2\gamma^2})^2(1+\Theta^2)^2}+\frac{a}{1+(\frac{\omega
a}{2\gamma^2})^2(1+\Theta^2)^2}\right]\!\!\Bigg\}.
\end{eqnarray}
The $\omega$-depending factor in the integrand is now a difference
of Lorentzians (bell-shaped functions of definite sign) having
different heights and widths determined by $\tau$ or $a$, but equal
areas. The first term in the brackets is negative, and can be
associated with LPM-like uniform suppression, whereas the second
term is positive, representing the anti-LPM enhancement, which is
now strictly positive and non-oscillatory. The shapes of
radiation-angle-integral spectra are illustrated in
Fig.~\ref{fig:Oscill-aver-t21} for same values of $\Sigma$ as in
Fig.~\ref{fig:Exact-oscill}, and (for the sake of illustration) a
moderately large ratio $\tau/a=3$. It may be observed that with the
increase of $\Sigma$, the shape of LPM-like suppression sharpens
towards $\omega\to 0$, although it does not tend to the thick-target
dependence $\sim\sqrt\omega$. Obviously, to get the $\sqrt\omega$
behavior, more than two scatterings need to be involved.

Finally, we note that in atomic matter, scales $\tau$ and $a$ should
vastly differ. To estimate $\tau$, it may suffice to note that
according to Migdal's calculation, typical $\omega$  at which LPM
effect develops are $\sim\frac{\gamma^2}{l_{\text{scat}}}$, where
\[
l_{\text{scat}}=\frac{l}{\Sigma^2(l)}\sim\frac{\alpha}{2\pi}
X_0\sim\frac1{8\pi Z^2\alpha^4}a_{\text{B}}
\]
is the range at which the angle of multiple scattering becomes
comparable with the radiation angle $\gamma^{-1}$. Hence, $\tau\sim
l_{\text{scat}}$, and owing to the presence of factor
$Z^{-2}\alpha^{-4}$, this range is at least by 4 orders of magnitude
greater than typical interatomic distance (a few units of
$a_{\text{B}}$), which must constitute the scale for $a$. Therefore,
anti-LPM enhancement in atomic matter must be virtually invisible
indeed, reconciling its principal existence with the conventional
theories of the LPM effect. Put differently, $|t_2-t_1|$ can be
effectively sent to zero, justifying the formal use of the
$\delta$-correlated model of multiple scattering employed by Migdal
\cite{Migdal}.

\section{Summary}\label{sec:summary}

In course of our study, we arrived at the following conclusions:
\begin{itemize}
\item
For scattering foils with macroscopic and fixed separation, the
LPM-like suppression of bremsstrahlung at lowest $\omega$ is
accompanied by an adjacent enhancement, which can be called the
anti-LPM effect. The latter effect may be associated with an
inclined-half-wavelength resonance [see
Eq.~(\ref{denom=1+Theta^2})], but not with the unit wavelength
resonance, as was previously suggested in \cite{Blankenbecler}. In
classical electrodynamics, LPM and anti-LPM effects are predicted to
have equal strength, in the sense that their $\omega$-integrated
effect is zero (see the end of Sec.~\ref{sec:2scat}). However, in a
random and weakly scattering media, like the ordinary atomic matter,
the anti-LPM enhancement is expected to smear over a broad spectral
interval, and thus become faint.

\item
The shape of the anti-LPM maximum depends on the strength of the
scattering in each plate (see Fig.~\ref{fig:Exact-oscill}). It
assumes universal forms for limits of strong or weak scattering: For
small scattering angles in both plates, the shape of the oscillatory
pattern in the spectrum is described by Eq.~(\ref{g-qq-def}).  For
large scattering angles in both plates, it is described by function
(\ref{E1}) (unifying different limiting cases described in
\cite{BK-structured}). For the mixed case when one plate scatters
strongly, and another one weakly, the spectrum shape obeys
Eq.~(\ref{g_ql-def}), where the dependence of  the interference term
on $\Sigma$ strictly saturates.

\item
At asymptotically large $\omega$, the dominant contribution to the
interference term stems from photon emission angles close to
$\bm{v}_2$, such that $|\bm{n}-\bm{v}_2|<\sqrt{\frac2{\omega
t_{21}}}<\frac1{\gamma}$. Therewith, the interference term scales as
$\propto\frac{\cos\Omega}{\Omega^2}$, with the coefficient
proportional to the product of $G$-amplitudes of radiation emitted
strictly parallel to $\bm{v}_2$ [see Eq.~(\ref{dI-largelambda})].
Granted this universality, the locations of \textit{secondary}
maxima and minima are virtually independent of the scattering
strength in the plates.
\end{itemize}

We have also established that if the rms scattering angle is
calculated with a proper account of Coulomb corrections, the
predictions of Gaussian averaging are pretty close to those of the
more precise Moli\`{e}re averaging, and practically comply with the
results of Zakharov \cite{Zakharov}, and Baier and Katkov
\cite{BK-structured}. For 2 scattering plates, the equation for
Moli\`{e}re averaging is basically as simple as that for Gaussian
averaging.

Finally, we note that satisfactory agreement of predictions of
classical electrodynamics with experiment was found. For future
experiment planning, it may be useful to refer to visibilities of
the interference minima and maxima at low and at large $\omega$,
computed in Sec.~\ref{sec:analysis} (Figs.~\ref{fig:V} and
\ref{fig:Vinfty}).

It may be added that prospects of observing anti-LPM effect in
radiation may be not restricted to artificial assemblies of plates,
provided one finds a case when $l_{\text{scat}}\sim a$, i.e., every
constituent of the matter can scatter the radiating projectile
relativistically. Conditions for such a situation may be sought, for
instance, in nuclear matter.

\subsection*{Acknowledgements}

This work was supported in part by UFFR Project No. 58/17.

\appendix

\section{Derivation of empirical formula (\ref{Highland-approx}) from Moli\`{e}re distribution}\label{appendix}

To clarify the success of the Gaussian approximation in conjunction
with interpolation (\ref{Highland-approx}) for the rms scattering
angle, consider the large-$\chi_c$ limit of Eq.~(\ref{exp-K1}):
\begin{eqnarray}\label{}
\left\langle F\right\rangle_{\text{M}}
&=&2\int_0^{\infty} d\rho \rho K_1^2(\rho)\left\{1-e^{-\frac{\chi^2_c}{\chi^2_1}\left[1-\rho\gamma\chi_1 K_1\left(\rho\gamma\chi_1\right)\right]}\right\}\nonumber\\
&\simeq&2\int_0^{\rho_0} \frac{d\rho}{\rho}\left\{1-e^{-\frac12\rho^2\gamma^2\chi^2_c\left(\ln\frac{2}{\gamma\chi_1\rho}+\frac12-\gamma_{\text{E}}\right)}\right\}\nonumber\\
&\,&+2\int_{\rho_0}^{\infty} d\rho \rho K_1^2(\rho)\nonumber\\
&\simeq&\int_0^{\rho_0^2} \frac{d\rho^2}{\rho^2}\left\{1-e^{-\frac14\rho^2\gamma^2\chi^2_c\left[\ln\left(\frac{\chi^2_c}{\chi^2_1}\ln\frac{\chi^2_c}{\chi^2_1}\right)+1-2\gamma_{\text{E}}\right]}\right\}\nonumber\\
&\,&+2\left(\ln\frac2{\rho_0}-\frac12-\gamma_{\text{E}}\right)\nonumber\\
&\simeq&\ln\left\{\gamma^2\chi^2_c\left[\ln\left(\frac{\chi^2_c}{\chi^2_1}\ln\frac{\chi^2_c}{\chi^2_1}\right)+1-2\gamma_{\text{E}}\right]\right\}-1-\gamma_{\text{E}}. \nonumber\\
\end{eqnarray}
Comparing it with Eq.~(\ref{Fsigma-la}), we establish the
correspondence
\begin{equation}\label{51}
\Sigma^2=\gamma^2\chi^2_c\left[\ln\left(\frac{\chi^2_c}{\chi^2_1}\ln\frac{\chi^2_c}{\chi^2_1}\right)+1-2\gamma_{\text{E}}\right].
\end{equation}
If we neglect here $1-2\gamma_{\text{E}}\approx -0.15$, and
substitute $\frac{\chi^2_c}{\chi^2_1}$ from Eq.~(\ref{chi2c/chi21}),
it casts as
\newpage
\begin{eqnarray}\label{}
\Sigma^2=\frac{\pi}{e^2\left(\ln\frac1{\gamma^2\chi_1^2}+\frac7{6}\right)}\frac{2l}{X_0}\qquad\qquad\qquad\qquad\qquad\qquad\nonumber\\
\times\left[\ln\frac1{\gamma^2\chi_1^2}+\ln\frac{\pi}{e^2}+\ln\frac{\ln\frac{\chi^2_c}{\chi^2_1}}{\ln\frac1{\gamma\chi_1}+\frac7{12}}+\ln\frac{l}{X_0}\right].\nonumber\\
\end{eqnarray}
Next, one can put  $\ln\frac{\pi}{e^2}\approx 6$,
$\ln\frac1{\gamma^2\chi_1^2}\approx 8\pm 2$,
$\ln\frac{\ln\frac{\chi^2_c}{\chi^2_1}}{\ln\frac1{\gamma\chi_1}+\frac7{12}}\sim
1$, and $\ln\frac{l}{X_0}$ can vary from $-7$ (if $l=10^{-3}X_0$) to
$-2$ (if $l=10^{-1}X_0$). As a result, $\Sigma$ can be approximated
by
\begin{equation}\label{}
\Sigma\approx
\frac{\mu}{m_e}\sqrt{\frac{2l}{X_0}}\sqrt{1+2c\ln\frac{l}{X_0}}\approx
\frac{\mu}{m_e}\sqrt{\frac{2l}{X_0}}\left(1+c\ln\frac{l}{X_0}\right),
\end{equation}
with $\mu\approx \sqrt{\frac{\pi}{e^2}
\frac{8+6+1}{8+7/6}}m_e\approx\text{13.6 MeV}$, and $c\approx
\frac{1}{2(8+6+1)}\approx 0.033$. Those numbers comply with the
coefficients in Eq.~(\ref{Highland-approx}), which we have thereby
derived ab initio. Parameter $\mu$ can also be compared with Rossi's
$E_s=\sqrt{\frac{4\pi}{e^2}}m_e=21.2$ MeV. Product
$\mu\left(1+c\ln\frac{l}{X_0}\right)$ becomes twice smaller than
$E_s$ at $l\sim 10^{-3}X_0$, which then exactly corresponds to
approximation (\ref{Blankenbecler-Q2def}). In general case, of
course, it is more reliable to use formula with an explicit $\ln
l/X_0$ dependence, like (\ref{Highland-approx}).

\end{document}